\documentclass[prodmode]{acmsmall} 

\usepackage[ruled]{algorithm2e}

\SetAlFnt{\small}
\SetAlCapFnt{\small}
\SetAlCapNameFnt{\small}
\SetAlCapHSkip{0pt}
\IncMargin{-\parindent}

\usepackage{times}

\usepackage{MnSymbol}
\usepackage{bm}
\usepackage{enumerate}
\usepackage{bbding}
\usepackage{wasysym}
\usepackage{color}
\usepackage{multirow}
\usepackage{rotating} 
\usepackage{graphicx}
\usepackage{wrapfig}
\usepackage{lscape}
\usepackage{pdflscape}
\usepackage{listings}
\usepackage[misc]{ifsym}
\usepackage{booktabs}

\usepackage[colorlinks=true,citecolor=blue,linkcolor=blue,urlcolor=blue]{hyperref}

\usepackage{longtable}

\usepackage{tikz}
\usetikzlibrary{calendar,fpu}
\tikzset{
    moon colour/.style={
        moon fill/.style={
            fill=#1
        }
    },
    sky colour/.style={
        sky draw/.style={
            draw=#1
        },
        sky fill/.style={
            fill=#1
        }
    },
    southern hemisphere/.style={
        rotate=180
    }
}

\makeatletter
\pgfcalendardatetojulian{2010-01-15}{\c@pgf@counta} 
\def\synodicmonth{29.530588853}
\newcommand{\moon}[2][]{%
    \edef\checkfordate{\noexpand\in@{-}{#2}}%
    \checkfordate%
    \ifin@%
        \pgfcalendardatetojulian{#2}{\c@pgf@countb}%
        \pgfkeys{/pgf/fpu=true,/pgf/fpu/output format=fixed}%
        \pgfmathsetmacro\dayssincenewmoon{\the\c@pgf@countb-\the\c@pgf@counta-(7/24+11/(24*60))}%
        \pgfmathsetmacro\lunarage{mod(\dayssincenewmoon,\synodicmonth)}
        \pgfkeys{/pgf/fpu=false}
    \else%
        \def\lunarage{#2}%
    \fi%
    \pgfmathsetmacro\leftside{ifthenelse(\lunarage<=\synodicmonth/2,cos(360*(\lunarage/\synodicmonth)),1)}%
    \pgfmathsetmacro\rightside{ifthenelse(\lunarage<=\synodicmonth/2,-1,-cos(360*(\lunarage/\synodicmonth))}%
    \tikz [moon colour=white,sky colour=black,#1]{
        \draw [moon fill, sky draw] (0,0) circle [radius=1ex];
        \draw [sky draw, sky fill] (0,1ex)
            arc (90:-90:\rightside ex and 1ex)
            arc (-90:90:\leftside ex and 1ex)
            -- cycle;
    }%
}

\newcommand{\fulladdress}{\CIRCLE}
\newcommand{\almostaddress}{\LEFTcircle}
\newcommand{\metioned}{\Circle}

\newcommand{\informal}{\Circle}
\newcommand{\semiformal}{\LEFTcircle}
\newcommand{\formal}{\CIRCLE}

\usepackage{pifont}
\newcommand{\cmark}{\ding{51}}%
\newcommand{\xmark}{\ding{55}}%

\newcommand{\todo}[1]{}

\newcommand{\sectprefix}{{Section}}
\newcommand{\subsectprefix}{{Subsection}}
\newcommand{\figprefix}{{Fig.}}

\newcommand{\tabprefix}{{Table}}

\newcommand{\tabincell}[2]{\begin{tabular}{@{}#1@{}}#2\end{tabular}}

\newcommand{\wenotknow}{?}

\usepackage{array}
\newcommand{\PreserveBackslash}[1]{\let\temp=\\#1\let\\=\temp}
\newcolumntype{C}[1]{>{\PreserveBackslash\centering}p{#1}}
\newcolumntype{R}[1]{>{\PreserveBackslash\raggedleft}p{#1}}
\newcolumntype{L}[1]{>{\PreserveBackslash\raggedright}p{#1}}

\usepackage[ruled]{algorithm2e}

\SetAlFnt{\small}
\SetAlCapFnt{\small}
\SetAlCapNameFnt{\small}
\SetAlCapHSkip{0pt}
\IncMargin{-\parindent}

\acmVolume{1}
\acmNumber{1}
\acmArticle{1}
\acmYear{2017}
\acmMonth{1}

\setcopyright{rightsretained}

\doi{0000001.0000001}

\issn{1234-56789}

\begin{document}

\markboth{Y. Zhao et al.}{High-Assurance Separation Kernels: A Survey on Formal Methods}

\title{High-Assurance Separation Kernels: A Survey on Formal Methods}

\author{Yongwang Zhao
\affil{Beihang University, China}
David San\'an
\affil{Nanyang Technological University, Singapore}
Fuyuan Zhang
\affil{Nanyang Technological University, Singapore}
Yang Liu
\affil{Nanyang Technological University, Singapore}
}

\begin{abstract}

Separation kernels provide temporal/spatial separation and controlled information flow to their hosted applications. 
They are introduced to decouple the analysis of applications in partitions from the analysis of the kernel itself. 
More than 20 implementations of separation kernels have been developed and widely applied in critical domains, e.g., avionics/aerospace, military/defense, and medical devices. Formal methods are mandated by the security/safety certification of separation kernels and have been carried out since this concept emerged. However, this field lacks a survey to systematically study, compare, and analyze related work. On the other hand, high-assurance separation kernels by formal methods still face big challenges.  
In this paper, an analytical framework is first proposed to clarify the functionalities, implementations, properties and standards, and formal methods application of separation kernels. Based on the proposed analytical framework, a taxonomy is designed according to formal methods application, functionalities, and properties of separation kernels. Research works in the literature are then categorized and overviewed by the taxonomy. 
In accordance with the analytical framework, a comprehensive analysis and discussion of related work are presented. Finally, four challenges and their possible technical directions for future research are identified, e.g. specification bottleneck, multicore and concurrency, and automation of full formal verification. 

\end{abstract}

\ccsdesc[500]{Software and its engineering~Operating systems}
\ccsdesc[500]{Software and its engineering~Formal methods}
\ccsdesc[300]{Computer systems organization~Real-time operating systems}
\ccsdesc[300]{Security and privacy~Formal methods and theory of security}

%
%

\terms{Design, Security, Verification}

\keywords{Separation Kernel, Formal Methods, Survey, Formal Specification, Formal Verification, Security, Safety}





\begin{bottomstuff}
This research is supported in part by the National Research Foundation, Prime Minister's Office, Singapore under its National Cybersecurity R\&D Program (Award No. NRF2014NCR-NCR001-30) and administered by the National Cybersecurity R\&D Directorate.
A minor part of this paper was presented in \cite{zhao16c}. 

Author's addresses: Y. Zhao (corresponding author), School of Computer Science and Engineering, Beihang University, Haidian District, Beijing 100191, P. R. China (Email: zhaoyw@buaa.edu.cn); D. San\'an, F. Zhang, Y. Liu, School of Computer Science and Engineering, Nanyang Technological University, 50 Nanyang Avenue, Singapore 639798 (Email:\{sanan,fuzh,yangliu\}@ntu.edu.sg). 
\end{bottomstuff}

\maketitle

\section{Introduction}
\label{sec:intro}

High-assurance systems require compelling evidences to show that their delivered services satisfy critical properties, e.g. security and safety \cite{mclean95}. If high-assurance systems fail to meet their critical requirements, it could result in security breaches, loss of lives, or significant property damage. 
Due to the criticality of such systems, it is highly desired that they are developed in a rigorous process. The avionics community has developed a set of guidelines for the rigorous development of safety-critical systems, e.g., DO-178B/C \cite{DO178B,DO178C}. Whilst the Common Criteria (CC) \cite{CC} provides guidelines for security-critical systems. 
In high-assurance systems, Trusted Computing Base (TCB) \cite{tcb85} is defined as: ``A small amount of software and hardware that security depends on and that we distinguish from a much larger amount that can misbehave without affecting security \cite{lampson92}''. 

The concept of \emph{separation kernel} is introduced \cite{Rushby81} to dissociate the kernel verification from the verification of trusted code belonging to separated components. 
The main purpose of separation kernels is to enforce the separation of all software components while reducing the
size of the TCB.
Security is carried out partly by separating physically system components, and partly by means of trusted functionality accomplished within some of those components being separated. 
The concept of separation kernel originates the Multiple Independent Levels of Security/Safety (MILS) \cite{Alves06} which is a high-assurance security/safety architecture based on separation \cite{Rushby81} and controlled information flow \cite{Denning76}. 
Separation kernels first came into use in the avionics domain, with the acceptance of Integrated Modular Avionics (IMA) \cite{Parr99} in this domain in the 1990s. 
A significant foundation of IMA is the separation of system resources into isolated computation spaces -- called \emph{partitions}. Separation kernels are adopted as \emph{partitioning kernels} \cite{pkpp}, which mainly concerns safety.

Separation kernels can be considered as a fundamental part of high-assurance systems. As a part of the TCB, separation kernels are small enough to allow formal verification of their correctness. 
The increasing evidences show successful applications of formal methods on software development, not only as theoretical research in the academy, but also deployed in industrial applications \cite{Woodcock09}. 
Traditionally, certified security is achieved by CC evaluation \cite{CC}, in which formal methods are mandated for highest assurance levels. It requires comprehensive security analysis using formal representations of the security model and functional specification as well as formal proofs of correspondence between them. In particular, the Separation Kernel Protection Profile (SKPP) \cite{SKPP07} is an instantiated profile of CC for separation kernels. Safety is usually governed by RTCA DO-178B \cite{DO178B} whose successor DO-178C \cite{DO178C} published in 2011 includes a technology supplement of formal methods.  

Due to the wide application of high-assurance systems, applying formal methods on separation kernels has not only been a hot research topic since the concept emerged, but also attracted industrial concerns. Although more than 20 implementations have been developed in industry or academia, and furthermore formal methods have been applied on some of them for the purpose of CC and DO-178B/C certification, high-assurance separation kernels still face challenges \cite{barhorst09,alan15,dmils14}. The approaches and techniques of formal methods for separation kernels are numerous, but the topic lacks a state of the art survey and a comprehensive taxonomy to ease the application of formal methods over them. It is therefore significant and urgent to have a thorough and comprehensive study on this topic to provide a useful reference for further research and industrial applications. 
To the best of our knowledge, our work is the first to systematically overview, categorize, analyze and discuss formal methods application on separation kernels. 

This paper aims at distilling the landscape in the field of formal methods application on separation kernels by studying, classifying, comparing, and analyzing related work for the purpose of figuring out challenges and potential research directions in future. Specifically, we present the following contributions in this paper:


(1) We propose an analytical framework to understand and classify related work. In the framework, we clarify a set of concepts related to separation kernels, define a reference architecture, compare implementations, study critical properties and related standards, and then identify an application schema of formal methods for separation kernels. The analytical framework is the foundation of this survey. 

(2) We propose a taxonomy of applying formal methods on separation kernels according to the analytical framework. The first level of the taxonomy is designed according to the application schema of formal methods. The lower levels are based on the reference architecture and critical properties. 
Then, we group together the related work according to the taxonomy. 

(3) We present a detailed analysis and discussion of the related work. We compare formal methods and certifications used in a comprehensive set of implementations. The importance of functionalities in the reference architecture is identified in formal specifications and models. Relations among critical properties are clarified. The verified properties, used approaches, and sizes of research works on formal verification are compared. Then, we give an overall comparison of them according to the taxonomy. 

(4) We discuss the challenges of applying formal methods on separation kernels and figure out potential research directions in this field. We identify four challenges, i.e., eliminating specification bottleneck, automating full formal verification, multicore and concurrency, and formal development and code generation. Then, we propose technical directions to address each challenge in future. 


Compared to our previous work \cite{zhao16c}, contributions (1), (3), (4) and the proposed taxonomy in contribution (2) in this paper are new. The detailed description of research works in \cite{zhao16c} is reorganized by the taxonomy and shortened to a brief overview of related work in contribution (2). The previous work is also extended by the research works of two new categories under the taxonomy. 
The rest of this paper is organized as follows. {\sectprefix} \ref{sec:analy_frm} presents the analytical framework. {\sectprefix} \ref{sec:tax} presents the taxonomy and overview of research works in the literature. In {\sectprefix} \ref{sec:analy_disc}, we analyze and discuss the research works through comprehensive comparisons. {\sectprefix} \ref{sec:chlg} identifies challenges and potential technical directions in this field. Finally, {\sectprefix} \ref{sec:concl} gives the conclusion of this survey. 


\section{Analytical Framework}
\label{sec:analy_frm}
In this section, we present an analytical framework for separation kernels. The framework is the foundation of the taxonomy, analysis and discussion in the next sections. First, we clarify a set of related concepts and propose a reference architecture for separation kernels in which common and optional components are identified. Second, we survey implementations from industry and academia. Third, 
we classify critical properties of separation kernels and survey related standards. Finally, we sketch out an application schema of formal methods for high-assurance separation kernels. 

\subsection{Concepts and Reference Architecture}
\label{ssec:cncpt_sk}

\begin{figure}
\centerline{\includegraphics[width=2.8in]{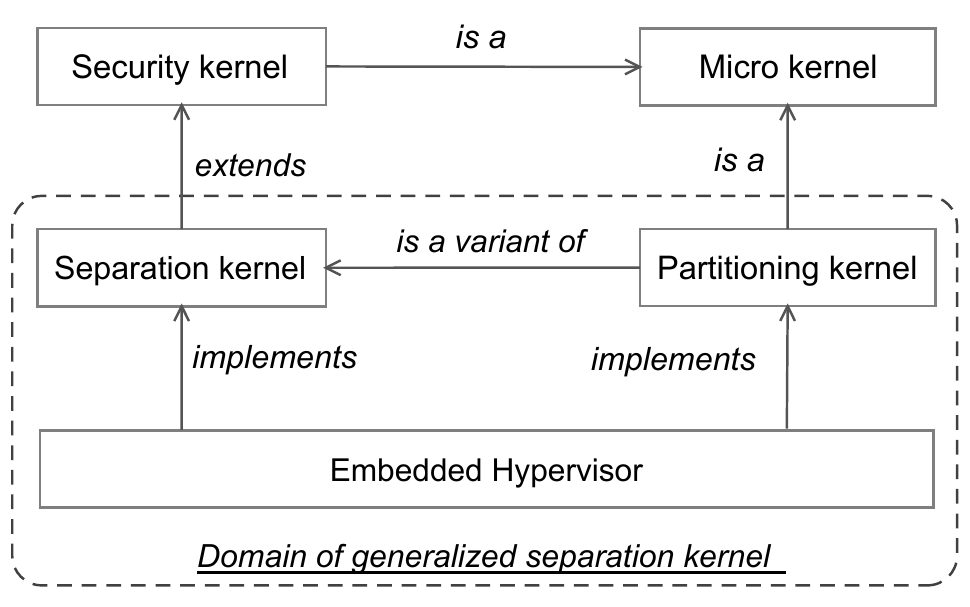}}
\caption{Relationship of Different Types of Kernels.}
\label{fig:kernels}
\end{figure}

We first clarify the relationship among concepts of security kernel, separation kernel, partitioning kernel, microkernel, and embedded hypervisor, which is shown in {\figprefix} \ref{fig:kernels}. 
The \emph{security kernel} \cite{Ames83} is the central part of systems to implement the basic security procedures for controlling access to system resources.
Security requirements of systems to be assured are specified as \emph{security policies}. A reference monitor controls the access of \emph{subjects} to \emph{resources} according to the policies.
Separation kernels extend security kernels with \emph{partitions} and map exported resources into partitions. Separation kernels enforce partitions to have spatial and temporal separation, and allow subjects belonging to partitions to cause flow to transfer information among them. 
The \emph{partitioning kernel} \cite{Rushby00,pkpp,Leiner07} is a variant of separation kernels in the domain of IMA and concerns safe separation largely based on an ARINC 653 \cite{ARINC653p0} style separation scheme. 
Partitioning kernels specialize and enhance the temporal and spatial separation with a static table-driven scheduling approach \cite{Ramam94} and static resource allocation for partitions.

Unlike traditional operating systems, separation kernels do not provide services such as device drivers and file systems, but a set of very specific functionalities to enforce security separation and information flow controls, in order to keep them small enough to allow formal verification of their correctness. 
The primary motivation of these kernels is also the one behind microkernels \cite{wulf74,Joch93,hoh04}. In terms of the source code size, these kernels are usually sizing less than 10,000 lines of code, which is the code scale of microkernels. 
On the other hand, with the rise of more powerful multiprocessor embedded systems, virtualization provides a promising technique to improve functionalities of high-assurance systems \cite{heiser08,agui12}. Embedded hypervisors are consequently used to implement security kernels (e.g. \cite{karger05,sailer05a,sailer05b}), separation kernels (e.g. \cite{xtratum,West16}), and partitioning kernels (e.g. \cite{vander10,han11,vander13}).

\begin{figure}
\centerline{\includegraphics[width=4.6in]{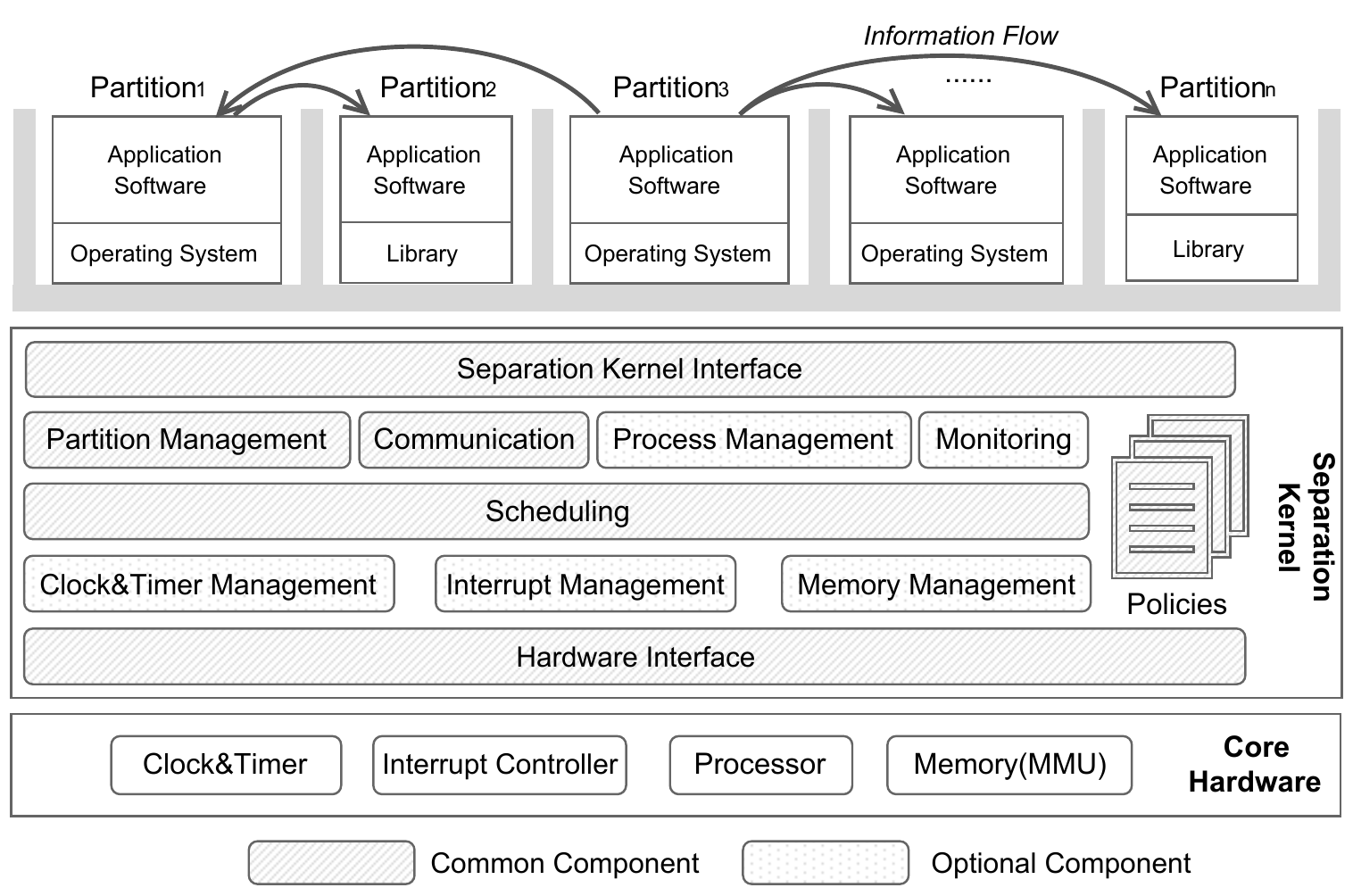}}
\caption{Reference Architecture of Separation Kernels.}
\label{fig:sk_ra}
\end{figure}

Due to the increasing complexity, scale, and mixed critical requirements of high-assurance systems, various techniques and approaches are integrated \cite{barhorst09} together. From now on, we use the term {\em separation kernel} to cover the concepts of security kernel, original separation kernel, partitioning kernel, and embedded hypervisor. Based on the landscape of separation kernels, we propose a reference architecture, as shown in {\figprefix} \ref{fig:sk_ra}, for separation kernels to provide functionalities to analyze research works. We classify the functionalities into common and optional components. \emph{Common components} represent a least set of functionalities to implement a separation kernel. \emph{Optional components} are usually supported for complex systems. 
Hypervisor-based separation kernels usually manage partitions (i.e., VMs) and leave process management to guest OSs. The communication mechanism supports inter- and intra-partition communication. Policies may be security, safety, real-time, and fault-tolerance policies, etc. The configuration for separation, such as memory separation configuration and scheduling windows for partitions, can also be considered in the policies. Management of hardware (e.g. clock, timer, interrupt, and memory) are necessary for hypervisor-based separation kernels. However, simple separation kernels manipulate the underlying hardware via \emph{hardware interface}.

\subsection{Separation Kernel Implementations}
\label{ssec:impl}

Due to the wide acceptance of separation kernels, many implementations including industrial products and academic prototypes have been developed in recent years. 
In {\tabprefix} \ref{tab:impls1}, we compare twenty implementations from industry and academia. The time line in the 3rd column shows the time they started and the time they stopped development of the separation kernels. The underlying instruction set architectures (ISA) and whether they support multi-core processors are surveyed in columns 4 and 5, respectively. We also survey the development languages, the line of the code (LOC), and whether they are open-source. 

\begin{table}%
\tbl{Comparison of Separation Kernel Implementations \label{tab:impls1}}{%
\begin{tabular}{|C{0.3cm}|L{3.8cm}|C{1.0cm}|C{2.2cm}|C{0.6cm}|C{1.2cm}|C{0.6cm}|C{0.7cm}|}
\hline
\textbf{No} & \textbf{Name} & \textbf{Timeline} & \textbf{ISA} & \tabincell{c}{\textbf{Multi} \\ \textbf{Core}}& \textbf{Language} & \textbf{LOC} & \tabincell{c}{\textbf{Open} \\ \textbf{Source}}
\\\hline \hline
\multicolumn{8}{|l|}{\textbf{Industrial Implementations}}
\\\hline 
1&PikeOS \cite{pikeos} & {\wenotknow} - now & PowerPC, x86, ARM, MIPS, SPARC& \cmark & C, ASM & $<$10k & \xmark
\\\hline
2&VxWorks 653 \cite{vxworks653} & {\wenotknow} - now & PowerPC & \cmark & C, ASM & {\wenotknow} & \xmark
\\\hline
3&VxWorks MILS \cite{vxworksmils}& {\wenotknow} - now & PowerPC & \cmark & C, ASM & {\wenotknow} & \xmark
\\\hline
4&INTEGRITY-178B \cite{integrity178b}& {\wenotknow} - now & ARM, x86, PowerPC, MIPS & \cmark & C, ASM & {\wenotknow} & \xmark
\\\hline
5&\tabincell{l}{INTEGRITY \\Multivisor \cite{integrityvisor}} & ? - now & x86, ARM, PowerPC & \cmark & C, ASM & {\wenotknow} & \xmark
\\\hline
6&LynxSecure \cite{lynxsec}& {\wenotknow} - now & x86 & \cmark & C, ASM & {\wenotknow} & \xmark
\\\hline
7&LynxOS-178 \cite{lynxos178}& {\wenotknow} - now & x86 PowerPC & \cmark & C, ASM & {\wenotknow} & \xmark
\\\hline
8&DDC-I Deos \cite{deos} & {\wenotknow} - now & x86, PowerPC, ARM, MIPS & \cmark & C, ASM & {\wenotknow} & \xmark
\\\hline
9&AAMP7$^a$ \cite{aamp7g} & 2001 - now & N/A & \xmark & & {\wenotknow} & \xmark
\\\hline
10&ED [\citeNP{Heitmeyer06};\citeyearNP{Heitmeyer08}]& 2006 - ? & {\wenotknow} & {\wenotknow} & C, ASM & $\approx$ 3k & \xmark
\\\hline
11&ARLX hypervisor \cite{arlx} & {\wenotknow} - now & x86, ARM & \xmark & C, ASM & {\wenotknow} & \cmark
\\\hline
\multicolumn{8}{|l|}{\textbf{Academic Implementations}}
\\\hline
12&seL4 [\citeNP{Murray12};\citeyearNP{Murray13}] & 2008 - now & ARM,x86 & \cmark & C, ASM & $\approx$ 9k  & \cmark
\\\hline
13&OKL4 Microvisor \cite{okl4mv} & 2009 - now & ARM & \cmark & C, ASM & {\wenotknow} & \cmark
\\\hline
14&XtratuM \cite{xtratum}& 2004 - now & SPARC, x86, PowerPC, ARM & \cmark & C, ASM & $\approx$ 9k  & \cmark
\\\hline
15&PROSPER \cite{prosper}& 2012 - now & ARM & \xmark & C, ASM & {\wenotknow} & \xmark
\\\hline
16&Xenon \cite{freit11} & 2011 - {\wenotknow} & x86, ARM, PowerPC & \cmark & C, ASM & {\wenotknow} & \xmark
\\\hline
17&Quest-V \cite{West16} & 2012 - now & x86 & \cmark & C, ASM & {\wenotknow}  & \xmark
\\\hline
18&Muen \cite{muen} & 2013 - now & x86 & \cmark & SPARK, ASM & $\approx$ 4k  & \cmark
\\\hline
19&POK \cite{pok} & 2009 - 2013 & PowerPC, SPARC, x86 & \xmark & C, ASM & $\approx$ 7k  & \cmark
\\\hline
20&AIR/AIR II \cite{air} & 2007 - 2011 & SPARC & {\wenotknow} & C, ASM & {\wenotknow}  & \xmark
\\\hline
\end{tabular}}
\begin{tabnote}%
\Note{{\wenotknow}} {means we do not find any literature to show the evidence.}
\vskip2pt
\tabnoteentry{$^a$}{This is a processor, a hardware implementation of separation kernel.}
\end{tabnote}%
\end{table}

By comparing these implementations, we have the following findings.
\begin{itemize}
\item Most of the separation kernels are still in use and developing. Very few open-source projects have stopped. 

\item In order to provide safety/security critical solutions, various ISAs are supported by separation kernels, in particular ARM, SPARC, and PowerPC. Multicore processors are increasingly deployed in safety/security critical systems to fulfil the demand of processing power in integrated systems. Therefore, multicore processors are supported by most of separation kernels regardless of in industry and academia. 

\item The LOC of separation kernels that we can find in the literature is less than ten thousand. Most of implementations adopt microkernels as the foundation and shift out the complex services into system partitions. For the sake of portability and efficiency, separation kernels in particulars are written in the C programming language embedded with pieces of ASM. Moreover, separation kernels in academia are usually delivered in open-source projects.  

\item With the trend of integrating applications on one computing platform (e.g., IMA), native interference provided by separation kernels is often not powerful for application development. The embedded hypervisor is currently a mainstream form of separation kernels in industry and academia. Virtual machine management provides a straightforward approach for the spatial separation of resources. Moreover, embedded hypervisors virtualize general-purpose operating systems (e.g., Linux) in partitions and permit the deployment of legacy applications.

\end{itemize}

\subsection{Critical Properties and Standards}
\label{ssec:cr_prop}

Traditionally, critical properties of high-assurance systems are safety, security, real-time, and fault-tolerance \cite{mclean95,rushby94}. Different from the classical categories of critical properties, \emph{NEAT} are well known properties considered in separation kernels, which stands for ``Non-bypassable, Evaluatable, Always invoked and Tamper'' proof \cite{Van05,pkpp}. 
However these intuitive concepts are not easy to formalize nor to provide direct proofs. Instead, separation kernels are normally verified by formally showing that they provide the right functionalities for MILS systems according to the following critical properties \cite{Alves06,Van05,Rushby00}, which is called \emph{DIDT} in this survey. 

\begin{itemize}
\item \textbf{Data Separation}: Also known as `Data Isolation'', each partition is deployed as a separated resource. Applications in one partition can neither modify applications and private data in other partitions nor control private devices and actuators in other partitions. 
\item \textbf{Information Flow Security}: Also known as ``Control of Information Flow'', information flow between partitions is defined from a source partition, which is authenticated, to a set of receivers as well authenticated; additionally, the source is authenticated to the receivers.
\item \textbf{Temporal Separation}: it allows partitions to share physical resources across different time periods. A resource is assigned to one component for a slice of time, then sanitized and assigned to another component. Services received from shared resources by applications in one partition cannot be affected by other partitions. 
\item \textbf{Damage Limitation}: damage is contained by restraining failures from propagating from one partition to others. 
\end{itemize}

The properties of data separation, information flow security, and damage limitation are all spatial properties. They are collectively called ``spatial separation'' properties. 
Data separation requires memory address spaces/data of one partition to be independent of any other partition in the system. Information flow security is a variation of data separation. Pure data separation is not pragmatic, therefore separation kernels define authorized channels between partitions to provide inter-partition communication. Pure data isolation is permitted to be violated only through these channels. Damage limitation is achieved by other three properties since the damage to applications in one partition are limited. 

Due to criticality of high-assurance systems, there are mandatory verification and validation (V\&V) activities in their design and analysis process to ensure that the systems fully meet their functional requirements. Several specifications have been created to standardize activities in V\&V processes by international organizations. 
CC \cite{CC}, which is also the international standard ISO/IEC 15408, and SKPP \cite{SKPP07} are usually applied to security of separation kernels. Although the SKPP was sunset in 2011, NSA still recommend separation kernels for security-critical systems. As for safety, \citeN{espo13} has summarized a set of well-known safety standards for high-assurance systems. Many of them have also been applied to separation kernels.
Another notable standard for separation kernels is the ARINC 653 standard \cite{ARINC653p0} which is a set of specifications to guide manufacturers in avionic application software towards maximum standardization. It aims at providing a standardized interface between separation kernels and application software, as well as the system functionalities of separation kernels.

In {\tabprefix} \ref{tab:props_stand}, we overview the traditional critical properties of high-assurance systems, their related standards, and whether DIDT properties contribute to improve the assurance of traditional critical properties. 

\begin{table}%
\tbl{Critical Properties and Standards \label{tab:props_stand}}{%
\begin{tabular}{|l|L{3.0cm}||C{1.3cm}|C{2.0cm}|C{1.3cm}|C{1.4cm}|}
\hline
\multirow{2}*{\textbf{Property}} & \multirow{2}*{\textbf{Standards}} & \multicolumn{3}{c|}{\textbf{Spatial Separation}} & \multirow{2}*{\tabincell{c}{\textbf{Temporal} \\ \textbf{Separation}}}
\\ \cline{3-5}
& & Data Separation & Information Flow Security & Damage Limitation & 
\\\hline
Safety & \tabincell{l}{DO-178B/C, ARINC 653, \\ IEC 61508, EN 50128} & \cmark & \xmark & \cmark & \cmark
\\\hline
Security & CC, SKPP & \cmark & \cmark & \cmark & \cmark 
\\\hline
Real-time & ARINC 653 & \xmark & \xmark & \xmark & \cmark 
\\\hline
Fault-tolerance & ARINC 653 & \cmark & \xmark & \cmark & \xmark
\\\hline
\end{tabular}}
\begin{tabnote}%
\Note{{\cmark}} {means a DIDT property contributes to improve the assurance of a traditional critical property.}
\end{tabnote}%
\end{table}

\subsection{Application Schema of Formal Methods}
\label{ssec:fm}
\begin{figure}
\centerline{\includegraphics[width=3.4in]{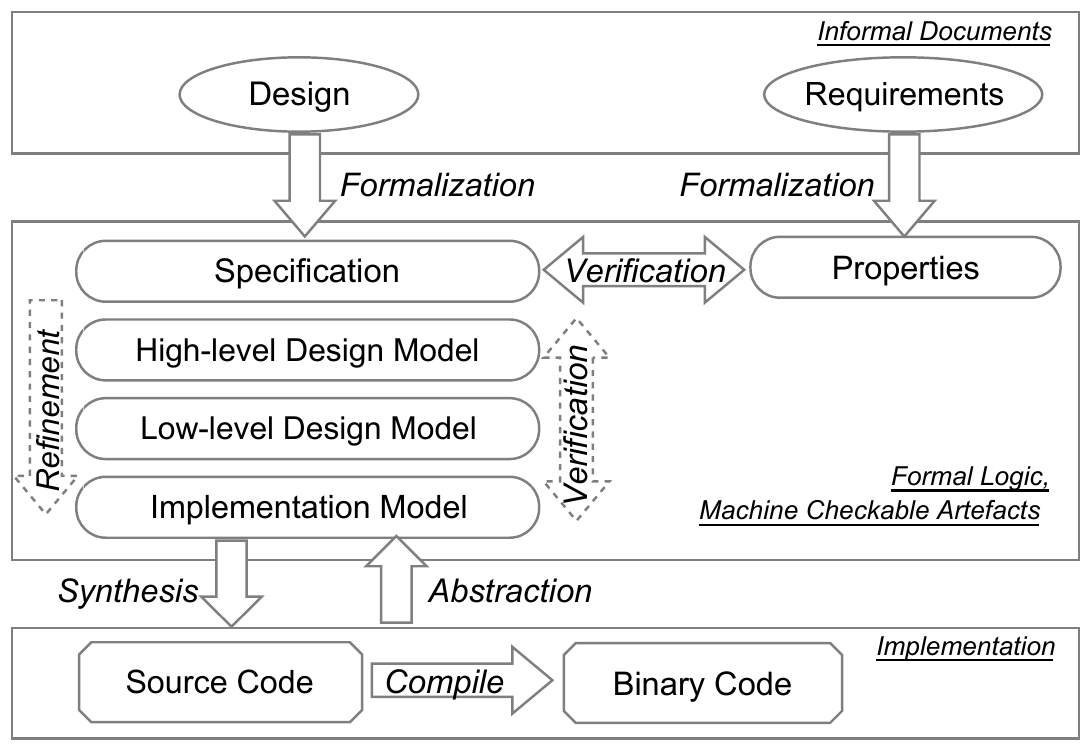}}
\caption{Application Schema of Formal Methods on Separation Kernels.}
\label{fig:fm_sw}
\end{figure}

In software engineering, formal methods provides a set of mathematically based techniques and tools to specify, develop, and verify software systems \cite{clarke96,bowen06}. We depict an application schema of formal methods on separation kernels in {\figprefix} \ref{fig:fm_sw}, in which the artefacts and techniques are identified. 

\begin{itemize}
\item Formal specification uses languages with a rigorous syntax and semantics to give a precise description of the system and its desired properties. Informal requirements may be translated into properties of the system specification. The system specification would further have formal description of system behavior, which is translated from the informal design. 
Formal specification can be used to validate the completeness and accuracy of the system requirements and to guide subsequent development activities.
Formal specification may be refined to high-level, low-level, and implementation models step by step and furthermore be used for formal synthesis of implementations. 

\item Formal verification is the act to ensure the correctness of intended systems with respect to a certain formal specification or property. One approach of formal verification is model checking, which systematically and exhaustively explores the mathematical model to check satisfaction of properties. Another one is theorem proving, whose first step is to generate a collection of proof obligations from the system and its specifications. The truth of the proof obligations implies the conformance of the system to its specification. The second step is to discharge the proof obligations in an interactive or automated manner. 

\item There are two approaches to formal verification of separation kernels at the implementation level:: theorem proving the implementation model by abstraction from source/binary code, and software model checking \cite{Jhala09}.
\end{itemize}

Many security and safety standards currently mandate the use of formal methods to certify correctness of separation kernels. The Common Criteria defines clear treatment of software artefacts for different evaluation levels, which is shown in {\tabprefix} \ref{tab:cc_eal}. 
The evaluation through CC defines Evaluation Assurance Levels (EAL) from EAL 1 to EAL 7 (formally verified, designed and tested). 
The EAL 7 mandates formal verification of the low-level design model using mathematical models and theorem proving.
As a specific profile of CC, SKPP mandates formal methods on separation kernels too.
DO-178C has a formal methods supplement (DO-333) to address formal methods to complement testing. 
IEC 61508 defines functional safety and methods for electronic systems. Certification of Safety Integrity Level (SIL) 4 in this standard highly recommend the use of formal methods.



\begin{table}%
\tbl{Common Criteria Evaluation Levels and Requirements of Formal Methods \label{tab:cc_eal}}{%
\begin{tabular}{|C{1.5cm}|C{1.8cm}|C{2.0cm}|C{1.8cm}|C{1.6cm}|c|}
\hline
\textbf{Common Criteria}&\textbf{Requirement}&\textbf{Functional Specification}&\textbf{High-Level Design}&\textbf{Low-Level Design}&\textbf{Implementation} 
\\\hline
EAL 1 -- 4 & \informal & \informal & \informal & \informal & \informal 
\\\hline
EAL 5 & \formal & \semiformal & \semiformal & \informal & \informal 
\\\hline
EAL 6 & \formal & \semiformal & \semiformal & \semiformal & \informal 
\\\hline
EAL 7 & \formal & \formal & \formal & \semiformal & \informal 
\\\hline
\end{tabular}}
\begin{tabnote}%
\Note{Formal methods level of software artefacts:} {\informal: informal; \semiformal: semiformal; \formal: formal}
\end{tabnote}%
\end{table}

\section{Taxonomy of Applying Formal Methods on Separation Kernels}
\label{sec:tax}
We first propose a taxonomy of applying formal methods on separation kernels in this section. The taxonomy is to group together related work that share common objectives and characteristics to yield clear category formation and easier comparative analysis. Then, we overview the related work using the proposed taxonomy. 

\subsection{Taxonomy}
\label{subsec:taxonomy}
\begin{figure}
\centerline{\includegraphics[width=5.5in]{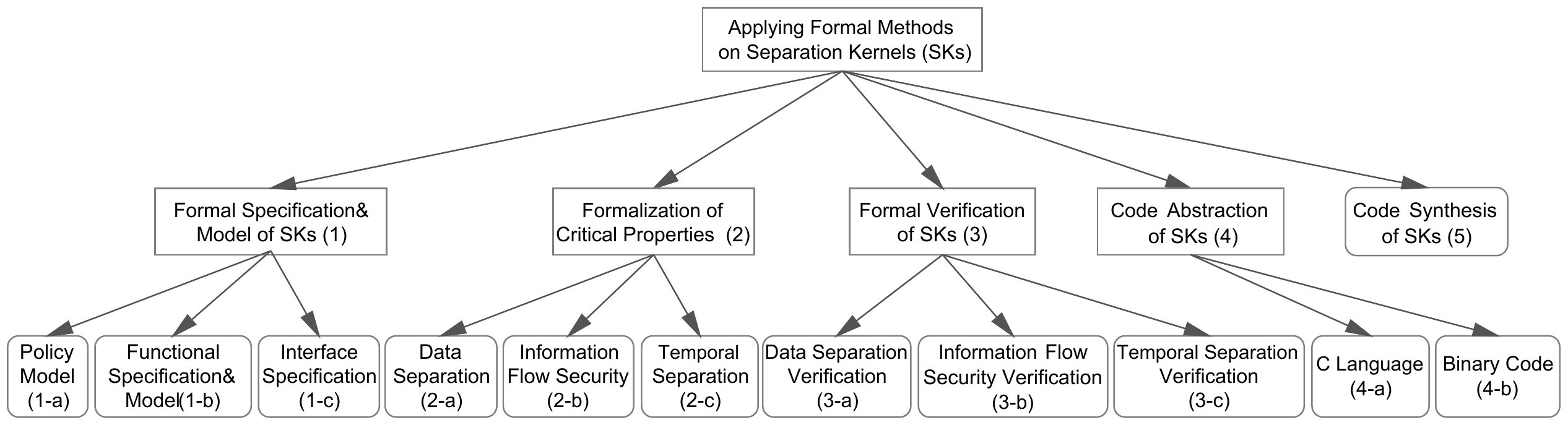}}
\caption{Proposed Taxonomy of Applying Formal Methods on Separation Kernel.}
\label{fig:taxonomy}
\end{figure}

The taxonomy is designed based on the analytical framework and is shown in {\figprefix} \ref{fig:taxonomy}. Level 0 is the root element. Level 1 of the taxonomy is designed according to the application schema of formal methods on separation kernels (see {\figprefix} \ref{fig:fm_sw}).  
Level 2 is designed considering the functionalities in the reference architecture (see {\figprefix} \ref{fig:sk_ra}), critical properties, and implementations. 
The subcategories of ``Formal Specification and Model of SKs'' are designed by considering functionalities in the reference architecture. 
The subcategories of ``Formalization of Critical Properties'' and ``Formal Verification of SKs'' are designed using the critical properties of separation kernels. Since the ``damage limitation'' property is enforced indirectly by the other three properties, there is no related work of formalization and formal verification of this property, and we omit it in our taxonomy. 
From the implementations of separation kernels (see {\tabprefix} \ref{tab:impls1}), we could see that they are almost developed using the C programming language. Thus, beside binary code we only consider code abstraction from the C language in the category ``Code Abstraction of SKs''. 
We discuss research works on each category in the following subsections.

\subsection{Formal Specification and Model of SKs (Category 1)}
\label{ssec:fun_spec}

This subsection overviews research works about formal description of functionalities of separation kernels. Except kernel interface and policies in {\figprefix} \ref{fig:sk_ra}, we group research works of other functionalities into the category ``functional specification and model''. 

\subsubsection{Policy Model (Category 1-a)}
\label{sssec:policy}
A formal policy is a kind of formal specification to describe what the system allows and prohibits. Formal policies of separation kernels actually define the security/safety requirements and can be categorized according to the critical properties. The policies are usually configured during system built-time and loaded during initialization of separation kernels. 

\emph{Data separation policy} defines strict data separation that does not allow data exchange between partitions. These policies include memory separation, device separation, etc. For instance, ARINC 653 defines a set of partitions and a static memory allocation policy for them \cite{ARINC653p13}. 

The \emph{inter-partition flow policy} (IPFP) \cite{Levin07} is a sort of information flow policy for separation kernels on MILS. Separation kernels map exported resources (e.g., communication objects) into partitions by a function $resource\_map: resource \rightarrow partition$. 
IPFP is expressed abstractly in a partition flow matrix $partition\_flow: partition \times partition \rightarrow mode$, whose entries indicate the mode of the flow. The mode indicates the direction of the flow, e.g. ``Write'' and ``Read''. 
Resources from a partition are addressed equivalently with respect to IPFP. One partition can be allowed to access all resources in another partition.
Another type of IPFP is port and channel based information flow used in ARINC 653. 
\emph{Partitioned information flow policy (PIFP)} \cite{Levin10} extends IPFP in SKPP with two different granularities of requirements: partitions and subjects/resources 
This abstraction allows subjects from a partition to have different access privileges to resources allocated in the same partition or even in a different partition. 

\emph{Fault policy} is a type of damage limitation policy. A typical fault policy for separation kernels is the health monitoring (HM) in ARINC 653. The HM reports and responds to hardware, kernel, and application faults and failures. ARINC 653 supports HM by providing a set of hierarchical HM configuration tables and application level error handlers. 
\emph{Scheduling policy} is a type of temporal separation policy. A typical scheduling policy for separation kernels is the partition time window configuration in ARINC 653. The scheduling specified in ARINC 653 is a two-level scheduling. The partition scheduling is a fixed, cycle based scheduling and is strictly deterministic over time. This cyclic scheduling consists of a major time frame (\emph{MTF}) that is split into partition time windows (\emph{PTW}). Each \emph{PTW} has an offset and a duration, which is associated to a given partition. 


\subsubsection{Functional Specification and Model (Category 1-b)}
\label{sssec:k_spec}
We overview a set of formal specifications and models of separation kernels here. Refinement is often applied to create concrete models from abstract specifications in a step-wise manner. We categorize research works according to specification languages used, i.e. system/software specification languages, formal languages in theorem provers, and architecture description languages. 


\paragraph{Using system/software specification languages}
A specification language is a high-level language other than a programming language for system analysis and design and to produce executable code. Many specification approaches use algebraic or model-theoretic structures to model systems step by step by refinement. 
In the following we describe related work using software specification languages, such as Z notation, B method, and Alley, to construct formal specifications of separation kernels.

\citeN{Craig07} concerns entirely with the specification, design, and refinement of operating system kernels in Z \cite{Abrial80}, one of which is a separation kernel. Refinement goes down to a level where source code in programming languages (e.g., C and Ada) can be extracted from Z specification. The specification and proofs are done by hand on paper. 
This work is upgraded in \cite{Velykis10} by  taking into account separation kernel requirements in \cite{Rushby81} and  SKPP \cite{SKPP07}. 
Craig's original specification is augmented using Z notation \cite{Wood96} mechanizing it using the Z/Eves theorem prover. As a consequent, syntax errors, missing invariants and new security properties to guarantee correct operations are found. 

The B Method \cite{abrial96} has been used to formally specify a secure partitioning kernel (SPK) in \cite{Andre09}. 
The high-level specification constitutes a complete architectural design of the system and is simulated and validated in ProB \cite{Leus03}. The PIFP policy is refined to a level from which C code can be automatically generated. Finally, an open source micro kernel, i.e., PREX, is adopted to integrate the PIFP implementation. 
Major functionalities of the OS-K separation kernel \cite{Kawamorita10}, such as partition management, inter-partition communication, access control, are also designed in the B method. Proof obligations are generated and checked using the B4free tool. Almost the whole totality of the 2,700 proof obligations comprising the verification are automatically proven using B4free.

Aiming at least privilege separation kernel (LPSK), \citeN{Phelps08} develop a formal security policy model and a top-level specification in Alloy \cite{jackson12}. They utilize the Alloy Analyzer to verify the consistency of the specification.
The top-level specification is a refinement of the PIFP policy model and uses state transitions to model two separation subsystems of LPSK, system initialization and the system during runtime.
In \cite{Martin00,Martin02}, three levels of abstraction and refinement are used to formally develop the MASK separation kernel in SPECWARE, which is an environment for formal specification and development. The abstract specification refines the MASK policies and concerns the communication among \emph{Cells} using \emph{strands}, which is a flow of instructions that are executed when a message is inserted into the strand of a cell. It is refined to the kernel specification primarily concerning the data structure. Finally, the bottom layer specification is manually translated into C source code. 
For the purpose of information flow security of the Xenon hypervisor \cite{mcdermott08}, 
\citeN{freit11} use \textsf{Circus} to formally model the hypercall interface behaviour of Xenon. \textsf{Circus} \cite{olive09} is a combination of Z, CSP and the refinement calculus. The whole model covers a subset of the hypercall interface and is over 4,500 pages of \textsf{Circus}.

\paragraph{Using theorem provers}
Theorem provers (e.g. Isabelle/HOL, Coq) generally have a small logical kernel, provide powerful expressive languages for specification, and support reasoning about high-order logic. They have been applied for formal verification of operating system kernels, such as seL4 \cite{Klein09,Klein14} and CertiKOS \cite{Gu15}. Inspired by successful application of theorem provers on general-purpose microkernels, they are adopted on kernels in recent years. 

The formal verification of the seL4 microkernel has been done using Isabelle/HOL \cite{Nipkow02}. 
The Isabelle/HOL specification of seL4 is extended in \cite{Murray13} to formally verify information flow security of seL4. In order to act as a separation kernel, seL4 is minimally extended by a static partition-based scheduler implementing a static round-robin scheduling between partitions, which are assigned fixe execution time slices. They also make small changes in the kernel APIs and add the security policy. 
Aiming at a precise model of PikeOS and a precise formulation of the PikeOS security policy, the EURO-MILS project \cite{euromils} creates a generic specification of separation kernels -- Controlled Interruptible Separation Kernel (CISK) \cite{Verb14} in Isabelle/HOL. This specification contains several facets that are useful to implement separation kernels, such as interrupts, context switches between domains, and control. 
The specification is rich in detail, making it suitable for formal verification of realistic and industrial systems.
\citeN{sanan14} construct in Isabelle/HOL the functional and security model of a generic partitioning separation microkernel from a reference specification based on European Space Agency's IMA for Space project \cite{Windsor11}. The specification uses ARINC 653 for the functional requirements and SKPP for the security requirements. Aiming at implementations, the specification covers hardware virtualization, CPU timer, and memory management too. \citeN{Zhao16} present a top-level specification of ARINC 653 compliant separation kernels in Isabelle/HOL, in which partition management, partition scheduling and communication services of ARINC 653 are considered. 

The Coq specification of CertiKOS in \cite{Gu15} is modified to disable all explicit inter-process communication and thus formed as a separation kernel without information flow among processes \cite{Costanzo16}. 
\citeN{AFBMRTO02} use the concept of virtual machine for separation and provide a formal model of a multi-partition systems (MPS) by ACL2 \cite{kaufm13}. Several different models of MPS are presented, including a two-partition system without communication between partitions and an $n$-partition system with restricted communication.

\paragraph{Using architecture description languages (ADLs)}
In general, ADLs concentrate on system level and are not fine-grained enough to formally specify separation kernels. However, formal models of separation kernels in ADLs could support model-driven development of applications. In \cite{Singhoff07}, AADL (Architecture Analysis and Design Language) and Cheddar \cite{Singhoff04} are applied to model an ARINC 653 hierarchical scheduler and to analyze the schedulability of applications represented by AADL specifications, respectively.

\subsubsection{Interface Specification (Category 1-c)}
\label{sssec:api}
The kernel interface defines operating system services provided to applications. Formalization of the kernel interface could support formally modelling and verification of application software on top of separation kernels. 

Formalization and verification of ARINC 653 has been conducted in recent years, such as a formal specification considering the architecture of ARINC 653 systems \cite{Oliveira12}, modeling ARINC 653 for IMA applications \cite{Wang11,Delan10}, and verification of application software on top of ARINC 653 \cite{de11}.
\citeN{zhao15} have formalized the system functionality and all of the 57 services specified in ARINC 653 Part 1 using Event-B \cite{Abrial07}. They use the refinement structure in Event-B to formalize ARINC 653 step by step and a semi-automatic translation from service requirements of ARINC 653 into the low level specification.

The formal API specification of PikeOS in Isabelle/HOL has been provided aiming at the certification of PikeOS up to CC EAL6 evaluation \cite{Klaus15}. Their specification is based on CISK \cite{Verb14}, which is instantiated to PikeOS API in detail. The formal API specification covers inter-partition communication, memory, file provider, port, and event.  



\subsection{Formalization of Critical Properties (Category 2)}
\label{ssec:sep_plcy}
Formal specification and model of separation kernels are verified with respect to critical properties. This subsection overviews the formal definition of critical properties and their sub-properties.

\subsubsection{Data Separation (Category 2-a)}
Data separation requires resources of a partition to be independent from resources from other partitions. Pure data separation is too strong since it does not permit communications among partitions. This property is relaxed in MASK \cite{Martin00,Martin02} and GWV \cite{Greve03}.
In the project of Mathematically Analyzed Separation Kernel (MASK) \cite{Martin00,Martin02}, communication between processes is regulated based on a separation policy, which is comprised of two separation axioms: a \emph{communication policy} and an anonymous policy. 
The communication policy states that if a cell $y$ is modified as the result of performing a step on a cell $x$, then there is an allowed communication between $x$ and $y$. 
The second policy requires that the execution of an action in a cell $x$ modifies the state of the cell $y$, then any modification in  $y$ has to depend only on $x$ and $y$. 
Based on the MASK data separation, \citeN{Greve03} propose the GWV property to model a separation kernel that enforces partitioning on applications running on mono-processors systems. The GWV property requires that the execution of a machine step modifying any arbitrary memory segment follows a mapping from the set of memory areas bound to the current partition and that are allowed to interact that memory segment. \citeN{Greve03} also define the \emph{exfiltration} and \emph{inflitration} properties for memory segments of partitions, which are special cases of the GWV property. The exfiltration and infiltration properties are similar to the communication policy and the second property of MASK respectively. 
\citeN{Heitmeyer08} apply the two axioms of MASK on the ED (Embedded device) separation kernel and define the \emph{no-exfiltration} and \emph{no-infiltration} properties for CC certification.

The GWV property has been accepted in industry \cite{integrity08,Greve04,Greve10} and formalized using the PVS theorem prover in \cite{Rushby04}. 
The original GWV is weakened by allowing to connect memory areas belonging to the same partition in \cite{Alves04}.
It is also extended by the concept of $subject$ and adding a restriction considering partition names in \cite{Tverdy11}. A subject is an element operating on memory areas of a partition.  
The GWV property has been applied in formal analysis for the INTEGRITY-178B separation kernel \cite{Richards10} and AAMP7 Microprocessor \cite{Greve04,Wilding10}.

Data separation of separation kernels at the hardware level is the separation of the system's memory. In \cite{Baumann11}, the memory separation of the PikeOS separation kernel is defined as ``All memory accesses in the kernel preserve an initial disjoint partitioning of memory, and obey a security policy where a thread is only allowed to access memory from its assigned partition.'' It is preserved by a set of assertions for function contracts. 

\subsubsection{Information Flow Security (Category 2-b)} 
\label{subsubsec:ifs}
Information flow security deals with the problem of preventing improper release and modification of information in complex systems.
Traditionally, language-based information flow security \cite{sabelfeld03} defines security policies of computer programs and ensures the data confidentiality by preventing information leakage from \emph{High} variables to \emph{Low} ones. 
Language-based information flow security is often not applicable for system-level security, because (1) in many cases it is impossible to classify \emph{High} and \emph{Low} variables; (2) data confidentiality is a weak property and is not enough for system-level security; and (3) language-based IFS is not able to deal with intransitive policies straightforwardly. Therefore, state-event based noninterference \cite{rushby92,von04}, which can deal with data confidentiality and secrecy of events together, is usually adopted in formal verification of separation kernels and microkernels \cite{Murray12}. We focus on state-event based properties in this paper. 


The concept of noninterference was introduced in \cite{Goguen82} for the purpose of the specification and analysis of security policies. The system is configured by a set of \emph{domains} and the allowed information flow between them are specified by an information flow policy $\rightsquigarrow$, such that $u \rightsquigarrow v$ if information is allowed to flow from the domain $u$ to the domain $v$. The intuitive meaning of noninterference is that a security domain $u$ cannot interfere with a domain $v$ if no action performed by $u$ can affect the observation of $v$ to the system.
Transitive noninterference is too strong and not able to model channel-control policies. Thus, intransitive noninterference is introduced in \cite{rushby92} as a declassification of transitive one. 
Based on noninterference in \cite{rushby92}, \citeN{von04} proposes new notions, \emph{nonleakage} and \emph{noninfluence}. Nonleakage is a state-event representation of language-based information flow security for arbitrary multi-domain policies. Noninfluence is the combination of nonleakage and intransitive noninterference. Intransitive noninterference and its new forms are usually chosen to formally verify information flow security of general purpose operating systems \cite{Murray12} and separation kernels \cite{Murray13}. Due to the scheduler in kernels, \citeN{Murray12} define special cases of nonleakage and noninfluence for operating systems. 
Properties of information flow security have been formally verified on seL4 \cite{Murray13}, PROSPER \cite{Dam13}, PikeOS \cite{Klaus15},  mCertiKOS \cite{Costanzo16}, and ARINC 653 \cite{Zhao16}. 

The standard proof of the noninterference property is discharged by examining a set of unwinding conditions \cite{rushby92} on individual execution steps of the system. The unwinding theorem states if the system is \emph{output consistent}, \emph{step consistent} and \emph{locally respects} the policy $\rightsquigarrow$, the system is secure for $\rightsquigarrow$. The three conditions are called \emph{unwinding conditions}. The unwinding theorem simplifies the security proofs by decomposing the global properties into unwinding conditions on each execution step.

\subsubsection{Temporal Separation (Category 2-c)}
Temporal separation usually includes sanitization/period processing and correct scheduling. \citeN{Heitmeyer08} define a sanitization property (called \emph{Temporal Separation}) on the ED separation kernel. The property ensures that the data areas of a partition are cleaned when the system is switched to process data in other partitions. As for period processing, time partitioning used in formal verification of DEOS scheduler \cite{Penix00,Penix05,Ha04,Cofer02} ensures that the access to CPU time budget by a partition cannot be affected by the execution of other partitions. Properties of correct scheduling are various according to different scheduling policies such as in \cite{Asberg11}. 


\subsection{Formal Verification of SKs (Category 3)}
\label{ssec:veri}
This subsection overviews research works about formal verification of separation kernels. We categorize the works by critical properties. 

\subsubsection{Data Separation Verification (Category 3-a)}

For the purpose of CC evaluation, Heitmeyer et al. \shortcite{Heitmeyer06,Heitmeyer08} provide a pragmatic solution to verify data separation of the ED separation kernel at the source code level. The kernel contains 3,000 lines of C and assembly code. To simplify the verification, the code is annotated in advance using Hoare and Floyd pre-post conditions. 
A top-level state machine is formally verified by data separation in TAME, which is a front end to the PVS theorem prover. Then the source code is partitioned and demonstrated to conform to the state machine by refinement. The effort of code verification is remarkably reduced since more than 90 percent of the source code is not corresponding to any behavior defined by the top-level state machine.

The AAMP7 microprocessors in Rockwell Collins is a hardware implementation of separation kernels. 
Their design is proven mathematically using the ACL2 theorem prover to achieve CC EAL 7 evaluation \cite{Greve04,Wilding10}. 
The \emph{intrinsic partitioning} in AAMP7 is an instantiation of the GWV property \cite{Greve03}. An abstract model meeting the GWV policy and a low-level model corresponding to the AAMP7 microcode are created, and the refinement between them is also proved.  

The INTEGRITY-178B separation kernel is formally analysed to obtain the EAL 6+ CC certification \cite{Richards10}. They adopt GWV \cite{Greve10} as the security policy and create three levels of specification, i.e., functional specification, high-level and low-level design, in ACL2. The functional specification is a formalization for the interfaces. The other two are semiformal representations of the system at different abstract levels. The low-level design has direct correspondence with the implementation, which simplifies the ``code-to-spec'' analysis during CC certification. 

\citeN{Tverdy11} presents a modular approach in Isabelle/HOL to the formal verification of the GWV property on the two layers of PikeOS. In the microkernel model, tasks and threads correspond to subjects and partitions in GWV respectively. A GWV \emph{segment} is instantiated as a physical memory address. 
They add ``partitions'' to the model of the separation kernel to separate tasks and physical address.
Memory separation of the PikeOS separation kernel has been formally verified on the source code level \cite{Baumann11} by breaking down high-level, non-functional requirements into functional properties of memory manager that can be presented as a set of assertions. 

\subsubsection{Information Flow Security Verification (Category 3-b)}

In the formal verification of the seL4 micro-kernel, to prove information flow security \citeN{Murray12} adopt the notions of \emph{nonleakage} and \emph{noninfluence} \cite{von04} and define their variations for OS kernels. The properties are formally verified on a revised specification of seL4 \cite{Murray13}. Because the properties are preserved by refinement, it is possible to first prove the information flow security property on the abstract model and then conclude that it holds for seL4's C source code due to the refinement relation. The verification applies to the total 8,830 lines of C code of the kernel implementation.

\citeN{Dam13} have formally verified information flow security of a simple ARM-based separation kernel -- PROSPER at the binary code level using HOL4. 
They construct the top level specification, which satisfies noninterference, and a real model, which consists of two partitions being executed on two independent machines targeting an ARMv7 processor, and connected by an explicit communication channel. They use the bisimulation proof method to show that user observable traces of the specification are the same as those of the real model. 
The approach avoids reliance on the correctness of a C compiler and can transparently verify C code mixing with assembly.

Explicit inter-process communication of mCertiKOS is disabled to form a strict separation kernel in which information flow among processes is not allowed. The noninterference property is verified in \cite{Costanzo16}. They use language-based information flow security and a well-designed observation function to express security at different abstract levels. A simulation preserves state indistinguishability between high and low levels. They develop a fully-formalized Coq proof to guarantee security of the assembly execution of mCertiKOS. 

Noninterference has also been formally verified on the PikeOS API specification \cite{Klaus15} and a top-level specification of ARINC 653 separation kernel \cite{Zhao16} using unwinding conditions. 

\subsubsection{Temporal Separation Verification (Category 3-c)}
Here, we discuss research works about formal verification of two-level scheduler which implements the partition scheduling in separation kernels. 

The Honeywell DEOS is a real-time operating system supporting flexible separation. Model checking and theorem proving approaches have been applied to the DEOS scheduler to check temporal separation \cite{Penix00,Penix05,Ha04}. 
A major part of C++ source code of the DEOS scheduler is first translated into Promela, which is the input language for the Spin model checker \cite{Penix00,Penix05}. Time partitioning is represented as a liveness property. The verification techniques are augmented in \cite{Cofer02} by verifying the absence of a livelock, which means that time is not elapsing in any cycle that does not contain a system tick event. Due to its size and complexity, state space explosion makes only possible to check one single configuration in each analysis. Thus, they turn to theorem proving approach and use PVS to analyze the scheduler \cite{Ha04}. To model the scheduler and the execution timeline in DEOS the authors use discrete time state-transition systems. Additionally, Time partitioning is expressed as a number of predicates that are proven to be true for any reachable states.

The Real-Time Specification for Java (RTSJ) is modified implementing a two-level scheduler. The first scheduling level is a priority scheduler to dispatch applications, while the second belongs to the applications \cite{Zerzelidis06b,Zerzelidis10}. The verification of this two level scheduler is carry out using time automata in UPPAAL. From a total of five verified properties, three of them concern the model correctness and others are liveness and deadlock free properties. 
In \cite{Asberg11}, a hierarchical scheduler for VxWorks has been modelled using task automata (timed automata with tasks) \cite{Fersman07} and automatically checked using the Times tool. They specify nine properties of the scheduler in TCTL (Timed Computation Tree Logic).

\subsection{Code Abstraction of SKs (Category 4)}

Formal verification of separation kernels down to their implementation or the binary code requires to provide a formal model for the semantics of the programming language or for the Instruction Set Architecture (ISA) of the target architecture, respectively. 
In this subsection, we overview the formal semantics and code abstraction of the C programming language and binary code. 

\subsubsection{Formal Semantics and Code Abstraction for C Language (Category 4-a)}

It is not until the end of the 1990's that semantics covering a subset of C, large enough to make the verification of complex and large programs possible, have appeared. \citeN{N98} brings in {\em Cholera} -- an operational semantics for C89 including the C type system. {\em Cholera} has been recently leveraged to construct the tools CParser and Autocorres \cite{GAK12}, which have been applied in the seL4 microkernel \cite{Klein09,Klein10} and separation kernel \cite{Murray13} to abstract the implementation model from the seL4 source code. 

\citeN{P01} develops a denotational semantics of C90, which is based on monads implemented in Haskell, and covers a large subset of C90. 
In the formal verification of the Nova hypervisor \cite{TWVPER08}, they provide a denotational semantics for C++ which includes all the C++ primitive datatypes. 
As part of the Verisoft project \cite{AHLSS08}, C0 which is a subset of the C language is formalized in Isabell/HOL. 
\citeN{BL09} develop Clight as part of the CompCert project \cite{Leroy09}. Clight accepts most of the C types and operators, although it does not support the use of control flow instruction {\tt goto} and blocks. 
\citeN{ER12} provide an executable semantics for C99 standard in the K-framework, which supports LTL model checking, and like the semantic model in \cite{P01}, it is not mechanized. They provide a semantic model for almost all of the C functionalities.
In the $CH_2O$ project, \citeN{KW15} provide a small step operational semantics and executable semantics model for C11 using the Coq theorem prover. The semantic model is non-deterministic covering almost the totality of the C standard. The executable semantics is used for validation purposes.

\subsubsection{Formal Semantics and Code Abstraction for Binary Code (Category 4-b)}
Related work in this area includes formalization of some of the most popular architectures such as Intel x86, ARM, and MIPS. Here we cover only those mechanized formal semantics that can be used in the binary code verification of separation kernels.

For the Intel architecture, \citeN{GHKG14} build in the model checker ACL2 an executable semantics for the x86-64 architecture, providing a framework able to both formally analyze and simulate non-deterministic machine code programs intended to run on 64 bits Intel processors. \citeN{SSNORBMA09} provide in HOL4 an axiomatic and operational semantics for the Intel multiprocessor architecture, including not only semantics for the set of instructions implemented by the architecture, but also a total order axiomatic semantic model of the memory, and machine registers. 

On ARM architectures, the work in \cite{FM10} covers ARM v7 including support for Thumb-2 instructions through a monadic encoding of the architecture operations. Validation is performed throughout random generation of instructions, and the execution of the instruction on a development board and in the semantic model. Recently, ARM v8 ISA is also modeled in \cite{Flur16}. The ARM v7 ISA model has been used in formal verification of seL4 \cite{Klein14} and PROPSPER \cite{Dam13} at the binary code level. 

Within the CompCert project \cite{Leroy09}, a subset of 90 instructions of the PowerPC ISA is modelled using the Coq theorem-prover. This semantics is extended using the HOL4 theorem-prover in \cite{AFIMSSN09} with an axiomatic memory model for multiprocessor. An x86
machine model derived from CompCert's model has been applied in formal verification of mCertiKOS \cite{Costanzo16}. 

It is worth to mentioning the L3 language, introduced in \cite{F15} aiming to support a generic framework for the specification of ISAs, and the reasoning on machine code programs. Through specifying a next-step function for a subset of instructions for a given ISA, and a definition of the state, the framework is able to generate high-level functions in HOL4 for machine code programs, and a set of theorems proving the correctness of the generated function w.r.t. the input machine code and the L3 specification for the ISA.


\subsection{Code Synthesis of SKs (Category 5)}
\label{ssec:sk_syn}
Formal synthesis \cite{Jullig93} translates formal, validated specifications into provably correct target code. 
Automated formal software synthesis gives a high degree of confidence that the generated code is correct with respect to the specification. Automatic code synthesis of operating systems can improve customizability \cite{Denys02} and optimize the performance at run-time \cite{Massa92}.
It is time consuming and error prone when manually porting or configuring the operating systems on different target architectures, and this issue can be addressed by automatic generation of application-specific operating systems \cite{Gauthier06}. 
But to the best of our knowledge, there are no research works on automatic code synthesis for separation kernels. The challenges are that the source code should be very efficient and usually embedded with assembly code. 
Actually, separation kernels in industries are always verified by the post-development approach, i.e., formal models are abstracted from the implementations of separation kernels and formally verified to provide the required proofs for critical properties. 

\section{Analysis and Discussion}
\label{sec:analy_disc}
In this section, we analyze and discuss related work from the perspective of implementations, formal specification and model, critical properties, formal verification, and code abstraction and synthesis. Then, we give an overall comparison. 

\subsection{Implementations}
We have surveyed twenty implementations of separation kernels from industry and academia in {\tabprefix} \ref{tab:impls1}.
Here, we compare their objectives, standard certifications/compliance, and formal methods applications in {\tabprefix} \ref{tab:impls2}. The ``objective'' column presents the critical properties that implementations concern. Although separation kernels contribute to improve fault-tolerance of systems, fault-tolerance is usually considered at system levels. Therefore, we do not compare this property in the table. 

\begin{table}%
\tbl{Comparison of Separation Kernel Implementations \label{tab:impls2}}{%
\begin{tabular}{|C{0.3cm}|L{4.8cm}|C{0.8cm}|C{0.6cm}|C{0.9cm}|C{2.4cm}|C{1.0cm}|}
\hline
\multirow{2}{*}{\textbf{No}} & \multirow{2}{*}{\textbf{Name}} & \multicolumn{3}{c|}{\textbf{Objective}} & \multirow{2}{*}{\tabincell{c}{\textbf{Certification} \\/\textbf{Compliance}}} & \multirow{2}{*}{\tabincell{c}{\textbf{Formal} \\ \textbf{Methods}}}
\\ \cline{3-5}
& & Security & Safety & Realtime & & 
\\\hline \hline
\multicolumn{7}{|l|}{\textbf{Industrial Implementations}}
\\\hline
1&PikeOS \cite{pikeos} & \cmark & \cmark & \cmark & \tabincell{c}{DO-178B Level B,\\IEC 61508 SIL 3,\\EN 50128 SIL 4, \\ARINC 653}& \cmark
\\\hline
2&VxWorks 653 \cite{vxworks653} & \xmark & \cmark & \cmark & \tabincell{c}{DO-178B/C Level A,\\ARINC 653} & {\wenotknow}
\\\hline
3&VxWorks MILS \cite{vxworksmils}& \cmark & \xmark & \xmark & \tabincell{c}{SKPP, CC,\\DO-178C Level A} & {\wenotknow}
\\\hline
4&INTEGRITY-178B \cite{integrity178b}& \cmark & \cmark & \cmark & \tabincell{c}{DO-178B Level A,\\ CC EAL 6+/SKPP, \\ ARINC 653}& \cmark
\\\hline
5&\tabincell{l}{INTEGRITY \\Multivisor \cite{integrityvisor}} & \cmark & \xmark & \xmark & {\wenotknow} & {\wenotknow}
\\\hline
6&LynxSecure \cite{lynxsec}& \cmark & \cmark & \cmark & \tabincell{c}{CC EAL 7,\\ DO-178B Level A} & {\wenotknow}
\\\hline
7&LynxOS-178 \cite{lynxos178}& \xmark & \cmark & \cmark & \tabincell{c}{DO-178B Level A,\\ARINC 653} & {\wenotknow}
\\\hline
8&DDC-I Deos \cite{deos} & \xmark & \cmark & \cmark & \tabincell{c}{DO-178B Level A,\\ARINC 653} & {\wenotknow}
\\\hline
9&AAMP7$^a$ \cite{aamp7g} & \cmark & \cmark & N/A & CC EAL 7& \cmark
\\\hline
10&ED [\citeNP{Heitmeyer06};\citeyearNP{Heitmeyer08}]& \cmark & \xmark & \xmark & CC & \cmark
\\\hline
11&ARLX hypervisor \cite{arlx} & \cmark & \cmark & \xmark & \tabincell{c}{DO-178B Level A,\\ MILS EAL,\\IEC 61508} & {\wenotknow}
\\\hline
\multicolumn{7}{|l|}{\textbf{Academic Implementations}}
\\\hline
12&seL4 [\citeNP{Murray12};\citeyearNP{Murray13}] & \cmark & \xmark & \xmark & \xmark & \cmark
\\\hline
13&OKL4 Microvisor \cite{okl4mv} & \cmark & \xmark & \xmark & \xmark & \cmark
\\\hline
14&XtratuM \cite{xtratum}& \cmark & \cmark & \xmark & ARINC 653 & \cmark
\\\hline
15&PROSPER \cite{prosper}& \cmark & \xmark & \xmark & \xmark & \cmark
\\\hline
16&Xenon \cite{freit11} & \cmark & \xmark & \xmark & \xmark & \cmark
\\\hline
17&Quest-V \cite{questv} & \cmark & \cmark & \xmark & \xmark & \xmark
\\\hline
18&Muen \cite{muen} & \cmark & \xmark & \xmark & \xmark & \xmark
\\\hline
19&POK \cite{pok} & \xmark & \cmark & \cmark & ARINC 653 & \xmark
\\\hline
20&AIR/AIR II \cite{air} & \xmark & \cmark & \cmark & ARINC 653 & \xmark
\\\hline
\end{tabular}}
\begin{tabnote}%
\Note{{\wenotknow}} {means we do not find any literature to show the evidence.}
\vskip2pt
\tabnoteentry{$^a$}{This is a processor, a hardware implementation of separation kernel.}
\end{tabnote}%
\end{table}

By comparing these implementations, we have the following findings. 
\begin{itemize}
\item Traditionally, two kinds of separation kernels have been used to assure safety and security of critical systems. For instance, VxWorks 653 was used to ensure safety-critical systems and VxWorks MILS to ensure security-critical systems. Nevertheless, a new direction in this field is to unify safety and security into a single separation kernel. For instance, recent separation micro-kernel implementations such as PikeOS and XtratuM are designed to support both solutions \cite{Zhao16}. 

\item The realtime property is mostly considered on separation kernels for safety-critical systems. Due to the integration of safety and security, this property has been considered with security-critical systems. 

\item Industrial implementations aim at highest assurance levels of different security/safety certification, in particular CC and DO-178B. Open-source/academic implementations have emerged in recent years. However, many of them do not have certification evidence now. Some of the open-source separation kernels are compliant with the ARINC 653 standard. 

\item From the aspect of formal methods application, formal specification and verification have been enforced on separation kernels in academia at source code and binary code levels. The objective of formal methods on industrial implementations is security/safety certification. 

\end{itemize}

\subsection{Formal Specification and Model}
We compare the research works of separation kernels on formal specification and model in {\tabprefix}s \ref{tab:sk_spec_comp1} and \ref{tab:sk_spec_comp2} in the ascending order of time. In {\tabprefix} \ref{tab:sk_spec_comp1}, we compare the formal languages they have used, the size of the specification, and whether refinement is used. In {\tabprefix} \ref{tab:sk_spec_comp2}, we compare the functionalities in the reference architecture that are formalized by the research works. We calculate a total score for each functionality to identify its importance in the formal specification and model of separation kernels.

\begin{table}
\tbl{Comparison of Separation Kernel Specification - Part 1 \label{tab:sk_spec_comp1}}{%
\begin{tabular}{|l|l|c|c|}
\hline
\textbf{Specification/Model}&\textbf{Formal Language}&\textbf{Size of Specification}&\textbf{Refinement}
\\\hline \hline
\tabincell{l}{MASK {[}\citeNP{Martin00};\citeyearNP{Martin02}{]}} & SPECWARE & {\wenotknow} & \cmark 
\\\hline
MPS \cite{AFBMRTO02}& ACL2 & $\approx$ 2500 LOC & \cmark 
\\\hline
Craig \cite{Craig07}&Z&$\approx$ 100 pages& \cmark 
\\\hline
\tabincell{l}{ARINC Scheduler \cite{Singhoff07}}& AADL & {\wenotknow} & \xmark 
\\\hline
LPSK \cite{Phelps08}& Alloy & {\wenotknow} & \cmark 
\\\hline
SPK \cite{Andre09}&Classical B&{\wenotknow}& \cmark 
\\\hline
OS-K \cite{Kawamorita10}&Classical B&{\wenotknow}& \cmark 
\\\hline
\tabincell{l}{Verified Software \cite{Velykis10}}&Z&$\approx$ 50 pages& \cmark 
\\\hline
Xenon \cite{freit11}& Circus & $\approx$ 4500 pages & \cmark 
\\\hline
seL4 \cite{Murray13} & Isabell/HOL & 4970 LOC & \cmark 
\\\hline
CISK \cite{Verb14}&Isabell/HOL&$\approx$ 500 LOC& \xmark 
\\\hline
XtratuM \cite{sanan14}&Isabell/HOL&$\approx$ 6000 LOC& \cmark 
\\\hline
ARINC 653 Standard \cite{zhao15}&Event-B& $\approx$ 2700 LOC & \cmark 
\\\hline
PikeOS API \cite{Klaus15}&Isabelle/HOL& $>$ 4000 LOC & \xmark 
\\\hline
ARINC 653 Separation Kernel \cite{Zhao16}&Isabelle/HOL& $\approx$ 1000 LOC & \xmark 
\\\hline
\end{tabular}}
\begin{tabnote}%
\Note{{\wenotknow}:}{means there is no evidence in the literature} 
\end{tabnote}%
\end{table}

\begin{table}
\tbl{Comparison of Separation Kernel Specification - Part 2 \label{tab:sk_spec_comp2}}{%
\begin{tabular}{|L{5.0cm}||C{0.15cm}|C{0.15cm}|C{0.15cm}|C{0.15cm}|C{0.15cm}|C{0.15cm}|C{0.15cm}|C{0.15cm}|C{0.15cm}|C{0.15cm}||C{0.15cm}|C{0.15cm}|C{0.15cm}|C{0.15cm}|}
\hline
\multirow{3}{*}{\textbf{Specification/Model}}&\multicolumn{10}{c||}{\textbf{Functionalities}} & \multicolumn{4}{c|}{\multirow{2}{*}{\textbf{Policies}}}
\\ \cline{2-11}
&\multicolumn{5}{c|}{\textbf{Common}} & \multicolumn{5}{c||}{\textbf{Optional}} & \multicolumn{4}{c|}{}
\\ \cline{2-15}

& \rotatebox{90}{\textbf{Kernel Interface}} & \rotatebox{90}{\textbf{Partition Management}} & \rotatebox{90}{\textbf{Communication}} &\rotatebox{90}{\textbf{Scheduling}} & \rotatebox{90}{\textbf{Hardware Interface}} & \rotatebox{90}{\textbf{Process Management}} & \rotatebox{90}{\textbf{Monitoring}} & \rotatebox{90}{\textbf{Clock\&Timer}} & \rotatebox{90}{\textbf{Interrupt}} & \rotatebox{90}{\textbf{Memory}} & \rotatebox{90}{\textbf{Data separation}} & \rotatebox{90}{\textbf{Inter-partition flow}} & \rotatebox{90}{\textbf{PIFP}}&\rotatebox{90}{\textbf{Fault}}

\\\hline \hline
\tabincell{l}{MASK \\ {[}\citeNP{Martin00};\citeyearNP{Martin02}{]}} & & $\Circle$ & $\CIRCLE$ & $\CIRCLE$ & & $\Circle$ & & & & & $\CIRCLE$ & & & 
\\\hline
MPS \cite{AFBMRTO02}& & $\Circle$ & $\LEFTcircle$ & & & $\Circle$ & & & & & $\LEFTcircle$&&&
\\\hline
Craig \cite{Craig07} &$\LEFTcircle$ & $\Circle$&$\CIRCLE$&$\CIRCLE$& $\LEFTcircle$& $\CIRCLE$&&&$\LEFTcircle$&$\CIRCLE$&&&&
\\\hline
\tabincell{l}{ARINC Scheduler\\ \cite{Singhoff07}}& & $\Circle$ & & $\CIRCLE$ & & $\Circle$ & & & & & &&&
\\\hline
LPSK \cite{Phelps08} & & $\Circle$ & & & & $\Circle$ & & & & $\Circle$ & &&$\CIRCLE$&
\\\hline
SPK \cite{Andre09} & $\LEFTcircle$&$\Circle$&$\LEFTcircle$&$\Circle$&&$\Circle$&&$\LEFTcircle$&&$\LEFTcircle$&&&$\CIRCLE$&
\\\hline
OS-K \cite{Kawamorita10} & $\LEFTcircle$&$\LEFTcircle$&$\LEFTcircle$&$\LEFTcircle$& & $\LEFTcircle$&&$\LEFTcircle$&&$\LEFTcircle$&&&&
\\\hline
\tabincell{l}{Verified Software\\\cite{Velykis10}} & $\LEFTcircle$&$\Circle$&$\LEFTcircle$&$\CIRCLE$&$\LEFTcircle$&$\CIRCLE$&&&$\LEFTcircle$&$\LEFTcircle$&&&$\CIRCLE$&
\\\hline
Xenon \cite{freit11} & $\LEFTcircle$ & $\Circle$ & $\LEFTcircle$ & $\LEFTcircle$ & $\LEFTcircle$ & $\Circle$ &  & & $\LEFTcircle$ & & &$\CIRCLE$&& 
\\\hline
seL4 \cite{Murray13}  & $\CIRCLE$ & $\Circle$ & $\CIRCLE$ & $\LEFTcircle$ & $\LEFTcircle$ & $\CIRCLE$ & & $\LEFTcircle$ & $\LEFTcircle$ & $\LEFTcircle$ & & $\CIRCLE$&&
\\\hline
CISK \cite{Verb14} & &$\Circle$&$\LEFTcircle$&$\Circle$&&&&&$\LEFTcircle$&&&&$\CIRCLE$&
\\\hline
XtratuM \cite{sanan14} & $\LEFTcircle$ &$\CIRCLE$&$\CIRCLE$&$\LEFTcircle$&$\LEFTcircle$&&$\LEFTcircle$&$\LEFTcircle$&$\Circle$&$\LEFTcircle$&&&$\CIRCLE$&
\\\hline
ARINC 653 Standard \cite{zhao15} & $\CIRCLE$&$\CIRCLE$&$\CIRCLE$&$\LEFTcircle$& & $\CIRCLE$&$\LEFTcircle$&$\LEFTcircle$&&&&&&$\LEFTcircle$
\\\hline
PikeOS API \cite{Klaus15} & $\LEFTcircle$&$\Circle$&$\LEFTcircle$&$\Circle$& & &&&$\LEFTcircle$&$\LEFTcircle$&&&$\CIRCLE$&
\\\hline
ARINC 653 Separation Kernel \cite{Zhao16} & $\LEFTcircle$&$\LEFTcircle$&$\LEFTcircle$&$\LEFTcircle$& & &&&&&&$\CIRCLE$&&
\\\hline
\multirow{2}{*}{\textbf{Total Score}}
&\multirow{2}{*}{24}&\multirow{2}{*}{21}&\multirow{2}{*}{31}&\multirow{2}{*}{27}&\multirow{2}{*}{10}& \multirow{2}{*}{20}&\multirow{2}{*}{4}&\multirow{2}{*}{10}&\multirow{2}{*}{13}&\multirow{2}{*}{16}& 5 & 9 & 18 & 2
\\ \cline{12-15}
&&&&&&&&&&&\multicolumn{4}{c|}{34}
\\\hline
\end{tabular}}
\begin{tabnote}%
\Note{$\CIRCLE$:}{has detailed specification (3 scores)} 
\Note{$\LEFTcircle$:}{has abstract specification (2 scores)} 
\Note{$\Circle$:}{only considers the concept (1 score)} 
\Note{}{The blank is that the specification does not cover the functionality (0 score)} 
\end{tabnote}%
\end{table}

A formal specification language has a mathematically defined syntax and semantics to give precise description of the artefacts used with formal methods. In the application of formal methods on separation kernels, numerous specification languages are used (see {\tabprefix} \ref{tab:sk_spec_comp1}), such as Classical B, Event-B, Z notation, Isabelle/HOL, ACL2, and model-driven architecture languages (e.g., AADL). 
Specification languages are often used for system analysis, requirement analysis, and systems design at a much higher level, where expressiveness and refinement \cite{RE08} are the major considerations for separation kernels. Specification languages used for separation kernels often support set theory and first-order logic as the fundamental data types. 
Refinement is often used to create concrete models from abstract specifications in a step-wise manner. 

On the other hand, for the purpose of formal verification at low level or source code level, specification languages used to specify separation kernels are focused on first-order or high-order logic languages, such as Isabelle/HOL, ACL2, PVS, and HOL4. Verification tools for these formalisms must have powerful engines for formal reasoning, supporting automatic theorem proving or providing proof assistants with a high degree of automation. \citeN{Wiedijk06} presents a detailed comparison of seventeen theorem provers and the ability of their formal notations.

We have proposed a reference architecture for separation kernels in {\figprefix} \ref{fig:sk_ra}, in which we classify the functionalities into common and optional components. In {\tabprefix} \ref{tab:sk_spec_comp2}, we count a total score for each functionality according to the level of abstraction at which they are formalized in research works. The importance of each functionality in the formal specification of separation kernels is thus shown by the total score. We divide the ``policies'' from ``functionalities'' in accordance with the taxonomy in subsection \ref{subsec:taxonomy}. 
From {\tabprefix} \ref{tab:sk_spec_comp2}, we could see that common components have higher scores than optional components. It is in accordance with the classification of common and optional components in the reference architecture. 
An exception is the ``hardware interface'' which is at low level and necessary in implementations. However, it is usually omitted in formal specifications at abstract level. 
Most of research works only consider the concept of ``partition'' and do not provide specification of ``partition management'', because they use the ``partition'' as a mechanism to separate resources and do not manage the life cycle of partitions. 
Although it is an optional component in the reference architecture, process management is often specified in research works because processes are importance resources in partitions. 
The PIFP policy is the most adopted policy of separation kernels in research works due to the fine-grained controls on partitions, resources, and subjects.

A notable observation is that different specification languages are used in the literature, but Isabell/HOL has become recently more popular than other formalisms. The first reason is its successful, large scale application in full formal verification of the seL4 microkernel at source code level. Second, there is a big community of experts working actively on its development, with frequent updates. Finally but not less important, it has a powerful development environment with many tools supporting automation. A detailed discussion about applying Isabelle/HOL in certification processes of separation kernels is in \cite{Blasum15}.

\subsection{Critical Properties}

\begin{figure}
\centerline{\includegraphics[width=5.0in]{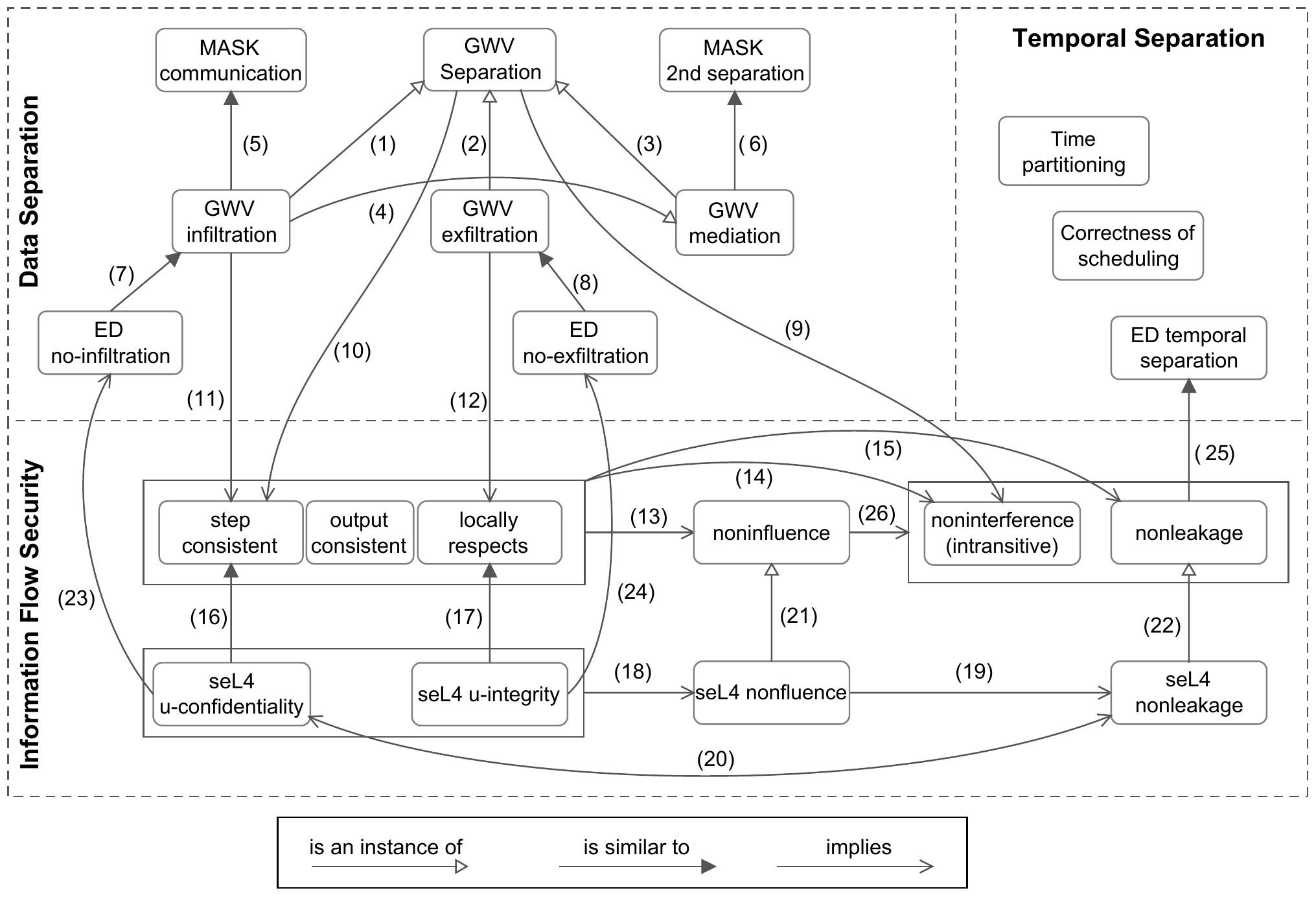}}
\caption{Relationship of Critical Properties.}
\label{fig:comp_props}
\end{figure}

In {\subsectprefix} \ref{ssec:sep_plcy}, we have presented a set of critical properties, their sub-properties, and their formal definition. We sketch out their relationship in {\figprefix} \ref{fig:comp_props}. The evidence of the relationship are from the literature as shown in {\tabprefix} \ref{tab:evd_prop}.

\begin{table}%
\tbl{Evidence of Relationship of Critical Properties \label{tab:evd_prop}}{%
\begin{tabular}{|l|l|}
\hline
\textbf{Literature}&\textbf{Labels of Relationship}
\\\hline \hline
\cite{Greve03} & (1 - 6)
\\\hline
\cite{Heitmeyer08,Heitmeyer06} & (7, 8)
\\\hline
\cite{Bond14} & (9, 10)
\\\hline
\cite{Alves04} & (9, 11, 12)
\\\hline
\cite{von04} & (13 - 15)
\\\hline
\cite{rushby92} & (14)
\\\hline
\cite{Murray12} & (16 - 20)
\\\hline
\cite{Murray13} & (23 - 25)
\\\hline
\cite{Zhao16} & (26)
\\\hline
\end{tabular}}
\end{table}

GWV and MASK are the two major groups of data separation properties. GWV is inspired by properties in MASK and the relationship of these two groups of properties is discussed in \cite{Greve03}. The infiltration, exfiltration, and mediation properties are actually instances of the GWV separation property. The first two are actually similar to the two properties of MASK. The properties of GWV are applied on the ED separation kernel and redefined as ED no-infiltration and ED no-exfiltration \cite{Heitmeyer08,Heitmeyer06}. 
Rushby's noninterference and its variants constitute the major group of information flow security. 
The definition and formal comparison of noninterference, nonleakage and noninfluence are studied in \cite{von04,Murray12,Zhao16}.
The three unwinding conditions (see {\subsectprefix} \ref{subsubsec:ifs}) imply noninterference by the unwinding theorem \cite{rushby92}. Noninfluence \cite{von04} is proposed based on noninterference and considers both data confidentiality and secrecy of events. It is a stronger property and implies noninterference and nonleakage. These properties have been instantiated in seL4 \cite{Murray12,Murray13} and in ARINC 653 separation kernels \cite{Zhao16} by extending the scheduler. Different definitions and formal comparison of noninterference are available in \cite{van2010}.

The GWV property proposed in Rockwell Collins is adopted in industry as the security policy for CC certifications, such as AAMP7 microprocessor \cite{Wilding10}, INTEGRITY-178B \cite{Richards10}, and PikeOS \cite{Tverdy11}. 
Meanwhile, noninterference is mostly applied in academia, such as in formal verification of seL4 \cite{Murray12,Murray13}, PROSPER \cite{Dam13}, and mCertiKOS \cite{Costanzo16}. 
A notable work is \cite{Bond14} in which they formally compare GWV and Rushby's noninterference and present a mapping between the elements of the two models. The conclusion is that GWV is stronger than Rushby's noninterference. A similar conclusion is in \cite{Alves04}, where they state that GWV is at least as strong as general noninterference and in addition it also provides intransitive noninterference.

Temporal separation has not attracted much attention in the literature and we therefore cannot find a large number of works focusing on the verification of temporal separation. It is however worth mentioning the work \cite{Penix00,Penix05,Cofer02} where time partitioning is verified in the DEOS kernel.


\subsection{Formal Verification}

In order to summarize formal verification of separation kernels, we compare the research works in {\tabprefix}s \ref{tab:sk_verify_comp1} and \ref{tab:sk_verify_comp2} focusing on the verification targets, verified properties and sub-properties, language used, sizes of specification and proofs, verification approaches and tools, and their cost. 
The verification targets are artefacts of formal methods in {\figprefix} \ref{fig:fm_sw}, i.e. specification, high-level model, low-level model, and implementation model of source code and binary code. In {\tabprefix} \ref{tab:sk_verify_comp1}, we refine the model into high-level design and low-level design models according to the levels of formal methods application in CC certification.

\begin{table}%
\tbl{Comparision of Separation Kernel Verification - Part 1\label{tab:sk_verify_comp1}}{%
\begin{tabular}{|L{3.8cm}|C{2.8cm}|C{2.8cm}|C{2.8cm}|}
\hline
\centering\textbf{Verified Kernel}& \textbf{Verification Target}& \textbf{Verified Properties} &\textbf{Verified Sub-properties} 
\\\hline \hline
ED \cite{Heitmeyer06,Heitmeyer08} & Implementation model (source code) & Data separation, Temporal separation & \tabincell{c}{No-infiltration, \\ No-exfiltration, \\ Kernel integrity, \\ Separation of control} 
\\\hline
AAMP7 \cite{Greve04,Wilding10}&Low-level model&Data separation & GWV
\\\hline
INTEGRITY-178B \cite{Richards10} &Low-level model&Data separation&GWV
\\\hline
PikeOS \cite{Baumann11} \cite{Tverdy11} \cite{Klaus15}&High-level model, Implementation model (source code) & Data separation, Information flow security &\tabincell{c}{Memory separation, \\ GWV, Noninterference}
\\\hline
seL4 \cite{Murray12,Murray13} & Implementation model (source code) & Information flow security &\tabincell{c}{Noninfluence, \\ Noninterference, \\ Nonleakage}
\\\hline
PROSPER \cite{Dam13} & Implementation model (binary code)&\tabincell{c}{Data separation, \\ Information flow security}&\tabincell{c}{No-infiltration, \\ No-exfiltration, \\ Noninterference}
\\\hline
XtratuM \cite{sanan14}&Low-level model& Information flow security & Noninterference 
\\\hline
mCertiKOS \cite{Costanzo16} & Implementation model (source code) & Information flow security & Noninterference 
\\\hline
ARINC 653 \cite{Zhao16} & Specification & Information flow security & \tabincell{c}{Noninfluence, \\ Noninterference, \\ Nonleakage}
\\\hline
DEOS \cite{Penix00,Penix05,Cofer02,Ha04}& Implementation model (source code) &Temporal separation&Time partitioning
\\\hline
A VxWorks scheduler \cite{Asberg11}&Low-level model&Temporal separation& \tabincell{c}{Correctness of Scheduling} 
\\\hline
RTSJ scheduler \cite{Zerzelidis06b,Zerzelidis10}&Low-level model&Temporal separation& \tabincell{c}{Correctness of Scheduling} 
\\\hline
\end{tabular}}
\end{table}

\begin{table}%
\tbl{Comparision of Separation Kernel Verification - Part 2 \label{tab:sk_verify_comp2}}{%
\begin{tabular}{|L{3.8cm}|C{1.5cm}|C{2.0cm}|C{1.2cm}|C{1.8cm}|C{1cm}|}
\hline
\centering\textbf{Verified Kernel}& \textbf{Formal Language} & \textbf{Size} &  \textbf{Verification Approach} & \textbf{Tools} & \textbf{Cost}
\\\hline \hline
ED \cite{Heitmeyer06,Heitmeyer08} & TAME & 368 LOC of TAME spec. & R, TP & TAME, PVS theorem prover & 11 weeks
\\\hline
AAMP7 \cite{Greve04,Wilding10} &ACL2& 3,000 LOC of ACL2 definitions & R, TP & ACL2 theorem prover & {\wenotknow}
\\\hline
INTEGRITY-178B \cite{Richards10} &ACL2& {\wenotknow} & R, TP &ACL2 theorem prover &  {\wenotknow}
\\\hline
PikeOS \cite{Baumann11} \cite{Tverdy11} \cite{Klaus15} & Annotated C code, Isabelle/HOL & {\wenotknow} & TP, MC& VCC, Isabelle proof assistant & {\wenotknow}
\\\hline
seL4 \cite{Murray12,Murray13} &Isabelle/HOL& 4,970 LOC of spec., 27,756 LoC of proof& R, TP & Isabelle proof assistant & 51 person-months
\\\hline
PROSPER \cite{Dam13} & HOL4& 21k LOC & R, TP &HOL proof assistant & {\wenotknow}
\\\hline
XtratuM \cite{sanan14}& Isabelle/HOL & 6,000 LOC of spec & R, TP &Isabelle proof assistant & 12 person-months
\\\hline
mCertiKOS \cite{Costanzo16} & Coq & 6,285 LOC of proof & R, TP & Coq proof assistant & {\wenotknow}
\\\hline
ARINC 653 \cite{Zhao16} & Isabelle/HOL & 1,000 LOC of spec, 7,000 LoC of proof & TP & Isabelle proof assistant & 8 person-months
\\\hline
DEOS \cite{Penix00,Penix05,Cofer02,Ha04}& Promela, PVS & 1,600 LOC of PVS &TP, MC & PVS theorem prover, SPIN & {\wenotknow}
\\\hline
A VxWorks scheduler \cite{Asberg11}& Task automata, TCTL& {\wenotknow} & MC &Times tool & {\wenotknow}
\\\hline
RTSJ scheduler \cite{Zerzelidis06b,Zerzelidis10}& Timed automata& {\wenotknow} & MC &UPPAAL & {\wenotknow}
\\\hline
\end{tabular}}
\begin{tabnote}%
\Note{Verification Approach:} {theorem proving (TP), model checking (MC), refinement (R)}
\Note{{\wenotknow}:}{means there is no evidence in the literature} 
\end{tabnote}%
\end{table}

The purpose of formal verification of separation kernels in industry is mainly safety/security certification, in particular CC security certification, such as INTEGRITY-178, PikeOS, AAMP7, and ED. The highest assurance level of CC certification (EAL 7) requires comprehensive security analysis using formal representations of the security model, functional specification, high-level design, and low-level design of separation kernels as well as formal proofs of correspondence among them. The implementation is not necessary for formal analysis. Therefore, it is possible to observe from {\tabprefix} \ref{tab:sk_verify_comp1} that formal verification of industrial separation kernels is often conducted on the low-level design model but not the implementation. However, in academia it often reaches the levels of source and binary code for the purpose of full formal verification. On the other hand, formal verification of separation kernels usually consider data separation and information flow security other than temporal separation since CC certification of separation kernels demands a security policy model of spatial separation. From the aspect of critical properties, formal verification in industry prefers data separation, in particular the GWV property, whilst Rushby's noninterference is prefered in academia. We find that in recent five years, research works of formal verification have mostly focused on the noninterference. 


Almost all of research works of formal verification on spatial separation have used theorem proving and refinement approaches.
The reasons are as follows. 
\begin{itemize}
\item The methodology of formal verification using theorem proving and refinement is compliant with CC EAL 7 certification. Security proof of separation kernels is produced by the methodology. However, model checking only produces the verification result, e.g., correct or counterexamples. 
\item Separation kernels for safety and security critical systems often requires formal verification on low-level design or even source code. 
Despite the relatively small size of separation kernels, the model checking technique does not scale well to verify such complex systems due to the state space explosion problem. However, the theorem proving approach is applicable and full verification of separation kernels is therefore possible.

\item Critical properties of separation kernels (e.g., GWV and information flow security) are difficult to be represented using temporal logic. A notable recent work is that noninterference can be classified as a sort of hyperproperties \cite{Clarkson10} and formulated by HyperLTL \cite{Clarkson14}. However, HyperLTL model checkers currently do not scale up to 1,000 states and are not applicable even at the abstract level of separation kernels. 
\end{itemize}

On the other hand, temporal separation verification often uses model checking rather than theorem proving. \emph{time} is hard to express using first order or high order logics, which are the mathematical artefacts used in theorem provers. However, it is possible to conveniently express \emph{time} using temporal logics, e.g., the timed automata in UPPAAL tools. A major obstacle of this approach is that the size and complexity of separation kernels limit the approach to analyze only one configuration at a time. Honeywell has addressed this issue and turned into using the PVS theorem prover to formally verify DEOS \cite{Ha04}.

From the aspect of the cost for formal verification, there are not many evidences in the literature. From the result of seL4 \cite{Klein09,Murray13}, we could see that enormous man power is often needed for formal verification of separation kernels reaching at the source code level. A possible approach to this issue is provided in \cite{Heitmeyer08}, where manual proof is enforced at abstract level and pre- and post-conditions annotated in the source code are used to automatically verify the conformance between the specification and the source code.

\subsection{Code Abstraction}

{\tabprefix} \ref{tab:sk_lang_semantics} summarises state of the art for mechanized formal semantics of the C language. The table shows the formal language used, the version of the C language the semantics formalizes, whether the formal semantics is executable, and what separation kernels they have been applied on, if any. We include a field indicating whether a semantics is executable or not since executing the semantics is a desirable property for simulation purposes. Similarly, {\tabprefix} \ref{tab:ISA_semantics} comprises state of the art ISAs formalization. The table shows the formal method used, the target architecture, whether it supports multicore, whether the semantics is executable, and what separation kernels they have been applied on, if any.
 
\begin{table}
	\tbl{Comparison of Formal Semantics of C \label{tab:sk_lang_semantics}}{%
		\begin{tabular}{|L{4.4cm}|C{1.8cm}|c|c|c|}
			\hline
			\textbf{Specification}&\textbf{Formal Language}&\textbf{C Subset}&\textbf{Executable}&\textbf{Applied Kernel}
			\\\hline \hline
			Cholera \cite{N98} & Isabelle/HOL & C89 & \xmark & CParser, seL4
			\\\hline
			\citeN{P01}& Haskell & C90 & \cmark &
			\\\hline
			\citeN{TWVPER08}&PVS& {\wenotknow} & \xmark & Nova Hypervisor
			\\\hline
			C0 \cite{AHLSS08}& Isabelle/HOL & C89 & \xmark &
			\\\hline
			Clight \cite{BL09}& Coq & C90 & \cmark &
			\\\hline
			\citeN{ER12}&K-framework&C99& \cmark &
			\\\hline
			$CH_2O$ \cite{KW15}&Coq&C11& \cmark &
			\\\hline
		\end{tabular}}
		\begin{tabnote}%
			\Note{{\wenotknow}:}{means there is no evidence in the literature} 
		\end{tabnote}%
	\end{table}

	\begin{table}
		\tbl{Comparison of Formal Semantics of ISA \label{tab:ISA_semantics}}{%
			\begin{tabular}{|L{3.8cm}|C{1.5cm}|C{1.4cm}|c|c|c|}
				\hline
				\textbf{Specification}&\textbf{Formal Language}&\textbf{ISA}&\textbf{Executable}&\textbf{Multicore}&\textbf{Applied Kernel}
				\\\hline \hline
				\citeN{GHKG14} & ACL2 & x86-64 & \cmark & \cmark & 
				\\\hline
				\citeN{SSNORBMA09}& HOL4 & x86-CC & \cmark & \cmark &
				\\\hline
				CertiKOS \cite{Costanzo16}& Coq & x86 & \cmark & \cmark & CertiKOS
				\\\hline
			     \citeN{FM10}&HOL4& ARMv7& \cmark & \xmark & seL4, PROSPER
				\\\hline
				\citeN{Flur16}&Lem& ARMv8& \cmark & \cmark &
				\\\hline
				CompCert \cite{Leroy09}& Coq & Power-PC & \cmark & \xmark &
				\\\hline
			\end{tabular}}
		\end{table}

Verification of separation kernels at source code and binary code level requires two fundamental tasks: to capture the language or ISA behaviour and to prove that the provided semantic model is correct. 
Due to the complexity of the C language, it is difficult to capture its whole semantics. Moreover, separation kernels are usually developed using C embedded with assembly. When it is formally verified at source code level, the kernel implementation is possible to be re-structured to make it compliant with the provided semantic model, such as in the seL4 microkernel verification \cite{Klein14}. This raises a new discussion about the validity of the new version of the kernel. 
However, although the modified kernel may not have the same behaviour as the original one, by the verification of functional correctness, which is carried out on the modified kernel, we obtain an implementation with the same functionality as the original kernel, and preserving the desirable set of safety and security properties.

Concerning the correctness of the C and ISA model, the technique most commonly used is the validation of the provided semantic model w.r.t. the programming language or ISA. A validation framework automatically executes single instructions, or sequences of them, in the semantic model and in the real architecture, and it compares the results of both execution to check whether they are correct. In case some mismatch is found, it is possible to refine the semantic model to correct a possible error on it. The validation of hundreds of thousands of instructions will provide enough confidence about the correctness of the semantic model.

\subsection{Overall Comparison}

We give an overall comparison of formal methods application on separation kernels in this subsection. 
We compare the related work on aspects of the target of formal methods application, development processes covered by the formal methods used according to CC, verification approaches, estimated EAL in CC according to {\tabprefix} \ref{tab:cc_eal}, and scale of formal methods application in {\tabprefix} \ref{tab:overall_comp}. 
In addition to the normative definitions of EALs, the CC standard defines the possibility of intermediate levels of security when a requirement is evaluated at a higher level than that required by the target level. The addition of the symbol ``+'' represents this kind of evaluation. 
Formal verification of ED, PikeOS, and PROSPER does not have low-level design. They either prove the conformance between the high-level design and the implementation or use software model checking to analyze the implementation. 

\begin{table}
\tbl{Overall Comparison \label{tab:overall_comp}}{%
\begin{tabular}{|L{4.6cm}|C{2.3cm}|C{0.10cm}|C{0.10cm}|C{0.10cm}|C{0.10cm}|C{0.10cm}|C{1.2cm}|C{0.6cm}|c|}
\hline
\centering\multirow{2}{*}{\textbf{Related Work}}& \multirow{2}{*}{\textbf{Target}} & \multicolumn{5}{c|}{\textbf{\tabincell{c}{Formal Methods \\ on CC Process}}} & \multirow{2}{*}{\textbf{Approach}} & \multirow{2}{*}{\textbf{EAL}} & \multirow{2}{*}{\textbf{Scale}} 
\\ \cline{3-7}
& & R & F & H & L & I & & & \rule{0pt}{0.3cm}
\\\hline \hline
GWV \cite{Greve03} & Properties & \cmark & \xmark & \xmark & \xmark & \xmark & TP & 4+ & +
\\\hline
Noninterference \cite{rushby92,von04,Murray12} & Properties & \cmark & \xmark & \xmark & \xmark & \xmark & TP & 4+ & ++
\\\hline
MASK \cite{Martin00,Martin02} & Low-level model & \cmark & \cmark & \cmark & \cmark & \xmark & R, TP, CG & 7 & {\wenotknow}
\\\hline
MPS \cite{AFBMRTO02} & High-level model & \cmark & \cmark & \cmark & \xmark & \xmark & R & 5+ & ++
\\\hline
An ARINC Scheduler \cite{Singhoff07} & Specification & \cmark & \cmark & \xmark & \xmark & \xmark & MC & 4+ & {\wenotknow}
\\\hline
Craig \cite{Craig07} & High-level model & \cmark & \cmark & \cmark & \xmark & \xmark & R, TP & 5+ & ++
\\\hline
LPSK \cite{Phelps08} & Specification & \cmark & \cmark & \xmark & \xmark & \xmark & TP & 4+ & {\wenotknow}
\\\hline
SPK \cite{Andre09} & High-level model & \cmark & \cmark & \cmark & \xmark & \xmark & R, TP, CG & 5+ & {\wenotknow}
\\\hline
OS-K \cite{Kawamorita10} & High-level model & \cmark & \cmark & \cmark & \xmark & \xmark & R, TP, MC & 5+ & {\wenotknow}
\\\hline
Verified Software \cite{Velykis10} & High-level model & \cmark & \cmark & \cmark & \xmark & \xmark & R, TP & 5+ & ++
\\\hline
Xenon \cite{freit11} & High-level model & \cmark & \cmark & \cmark & \xmark & \xmark & R, TP & 5+ & +++
\\\hline
CISK \cite{Verb14} & High-level model & \cmark & \cmark & \cmark & \xmark & \xmark & TP & 5+ & ++
\\\hline
ARINC 653 Standard \cite{zhao15} & High-level model & \cmark & \cmark & \cmark & \xmark & \xmark & R, TP & 5+ & ++
\\\hline

ED \cite{Heitmeyer06,Heitmeyer08} & Implementation model (source code) & \cmark & \cmark & \cmark & \xmark & \cmark & R, TP & 7 & ++ $^a$
\\\hline
AAMP7 \cite{Greve04,Wilding10} & Low-level model & \cmark & \cmark & \cmark  & \cmark & \xmark & R, TP & 7 & ++ $^b$
\\\hline
INTEGRITY-178B \cite{Richards10} & Low-level model & \cmark & \cmark & \cmark & \cmark & \xmark & R, TP & 6+ & {\wenotknow}
\\\hline
PikeOS \cite{Baumann11,Tverdy11,Klaus15} & High-level model, Implementation model (source code) & \cmark & \cmark & \cmark  & \xmark & \cmark & TP, MC & 6+ & ++ $^c$
\\\hline
seL4 \cite{Murray13} & Implementation model (source code) & \cmark & \cmark & \cmark & \cmark & \cmark & R, TP, CA & 7+ & ++++
\\\hline
PROSPER \cite{Dam13} & Implementation model (binary code) & \cmark & \cmark & \cmark & \xmark & \cmark & R, TP, CA & 7+ & +++
\\\hline
XtratuM \cite{sanan14} & High-level model & \cmark & \cmark & \cmark & \xmark & \xmark & R, TP & 5+ & ++
\\\hline
mCertiKOS \cite{Costanzo16} & Implementation model (source code) & \cmark & \cmark & \cmark & \cmark & \cmark & R, TP, CA & 7+ & +++
\\\hline
ARINC 653 \cite{Zhao16} & Specification & \cmark & \cmark & \xmark & \xmark & \xmark & TP & 5 & ++
\\\hline
DEOS \cite{Penix00,Penix05,Ha04} & Implementation model (source code) & \cmark & \cmark & \cmark & \xmark & \cmark & TP, MC & 7 & {\wenotknow}
\\\hline
A VxWorks scheduler \cite{Asberg11} & High-level model & \cmark & \cmark & \cmark & \xmark & \xmark & MC & 5+ & {\wenotknow}
\\\hline
RTSJ scheduler \cite{Zerzelidis06b,Zerzelidis10} & High-level model & \cmark & \cmark & \cmark & \xmark & \xmark & MC & 5+ & {\wenotknow}
\\\hline
\end{tabular}}
\begin{tabnote}%
\Note{Approach:} {Refinement (R), Theorem Proving (TP), Model Checking (MC), Code Abstraction (CA), Code Generation (CG)}
\Note{Formal methods on CC process:} {Requirement (R), Functional specification (F), High-level design (H), Low-level design (L), Implementation (I)}
\Note{Scale:} {+ (\textless 1k LOC), ++ (1k $\sim$ 10k LOC), +++ (10k $\sim$ 100k LOC), ++++ (\textgreater 100k LOC). The LOC includes the specification and proof}
\tabnoteentry{$^{a,b,c}$}{ Only LOC of specification is available.}
\end{tabnote}%
\end{table}

The highest assurance level of CC (EAL 7) requires formal methods application on the low-level design but not on the implementation. Aiming at security/safety of separation kernels as far as possible, a few research works have provided formal proof of refinement between the low-level design and the implementation, such as seL4 \cite{Murray13} and mCertiKOS \cite{Costanzo16}. Targeting at the source code by formal methods always means that they are applied on the implementation, which has overstepped the demand of EAL 7 in CC and estimated as EAL 7+ in {\tabprefix} \ref{tab:overall_comp}. Compared to the high cost of CC certification, formal verification on implementation is a low-cost way to provide more assurance of separation kernels as stated in \cite{Klein09}. 

Few work provides an estimation of time and cost of formal methods application on separation kernels. Thus, we cannot clearly compare them. A notable viewpoint is that the industry rules-of-thumb for CC EAL 6 certification is of $\mathdollar$1k/LOC, although it provides less assurance than formal verification \cite{Klein14}. 

\section{Challenges and Future Directions}
\label{sec:chlg}

We now discuss the remaining challenges in formal methods application on separation kernels and possible research directions that may be taken by future work. 

\subsection{Eliminating Specification Bottleneck}

In formal methods, formal specification is a bottleneck in functional verification \cite{Beck14}. Therefore, simpler verification methods are often used in practice including (1) lightweight verification methods for finding bugs, (2) combining verification and testing, and (3) verifying generic and uniform properties. 
Due to high assurance of separation kernels and formal methods mandated by certification, the first method is obviously not sufficient. The second is always used in practice. The third is actually suitable for separation kernels. 
Using generic or uniform specifications can reduce the cost to create requirement specifications. Although lightweight properties, such as buffer overflows and null-pointer exceptions, are feasible in many cases, formal verification of relational properties, e.g. noninterference, is inevitable for separation kernels. The challenges and possible directions to eliminate specification bottleneck are shown as follows.

(1) \emph{Properties of temporal separation}. The GWV and noninterference are the major properties for data separation and information flow security that have been widely applied in industry and academia. However, properties of temporal separation have not been thoroughly studied in the literature. A set of properties to clarify temporal separation are highly desired for high-assurance separation kernels. 

(2) \emph{Formal relations among properties}. We have figured out some formal relations in {\figprefix} \ref{fig:comp_props}. Others are not explored. In particular, shared resources among partitions can affect the scheduling in separation kernels. But the relationship between spatial separation and temporal separation has not been studied in the literature and is not clear yet. On the other hand, there does not exist a precise and global framework for the relationship of critical properties of separation kernels and it deserves further study. 

(3) \emph{Generic formal specification}. For the purpose of the formal development, safety/security certification, and the study of formal relations between critical properties, it is highly necessary in the future to create a generic specification of separation kernels. This specification can be used to develop implementations using refinement and be revised gradually, and thus significantly alleviate the bottleneck. It has been attempted in the EURO-MILS project to deliver a generic specification for separation kernels \cite{Verb14}.

(4) \emph{Reusability of formal specification}. Formal specification is a foundation for formal verification. Furthermore, it can also be applied to development, integration, and management of systems deployed on separation kernels \cite{Zhao16b}. A direction is to integrate domain knowledge into formal specification of separation kernels (e.g., \cite{AitAmeur16,Zhao16b}) to improve its reusability. 

(5) \emph{Flexibility of formal specification and proof}. Although reusability of the specification partially relies on the formal notation used and its supported tool, a well designed specification can evidently improve it. On the other hand, proofs should address how to deal with changes of formal model due to upgrading of separation kernels. Since re-verification is usually expensive for separation kernels, the proof change should be as small as possible when the uniform specification and design models are tailored or extended in real applications. A reusable design of the specification and its proof is a challenge for separation kernels.


\subsection{Automating Full Formal Verification}
Full formal verification of systems means that the verification is enforced not only on the specification but also covers all the source code and even the binary code with machine checkable proof. Formal verification at the implementation level can significantly improve the assurance of systems than other approaches, such as applying formal specification or lightweight properties over higher-level models \cite{Andronick12}.
Full formal verification of programs had rarely been conducted and was often considered to be highly expensive \cite{Hall90} before successful practices of seL4 \cite{Klein09}, CompCert \cite{Leroy09}, and CertiKOS \cite{Gu15}. 

Full formal verification at the source code level is necessarily based on a set of assumptions, such as the correctness of the hardware and the compiler \cite{Klein09}. Whilst, formal verification at the binary code level overcomes assumptions on the correctness of the compiler. Full correctness of separation kernels by formal verification could be assured by a formal pervasive verification approach covering the hardware, compiler, and kernel itself exactly as proposed in the Verisoft XT project \cite{Hille08}. 
A major obstacle of this objective is that full formal verification of operating system kernels is usually manpower intensive, e.g., 20 person-years are invested in formal verification of seL4. We summarize a set of challenges and potential directions in automating full formal verification of separation kernels to alleviate enormous efforts as follows.

(1) \emph{Automatic verification of critical properties}. As shown in {\tabprefix} \ref{tab:sk_verify_comp2}, existing works usually apply theorem proving to verify spatial separation of separation kernels. Automatic approaches at specification and design levels can enormously alleviate manual efforts and deserves further study. 

(2) \emph{Automatic refinement checking and property preservation}. A promising way to the correctness of low-level models is refinement. However, from seL4 we could see that it is often a time-consuming work to find and prove the refinement relation between two levels of specification \cite{Klein09}. Automatic refinement checking is thus worth considering in formal verification of separation kernels. \citeN{Zhao16b} have illustrated high degree of automatic refinement checking using Event-B. Second, it is critical that properties could be preserved during refinement. Refinement preservation of information flow security has been discussed in \cite{van12}. For separation kernels, refinement preservation of critical properties needs systematical study. 

(3) \emph{Proof generation during automatic verification}. Traditional model checking approaches produce the verification result directly. For the purpose of safety and security certification, it is necessary that automatic approaches generate proofs for the correctness. 

(4) \emph{Full formal verification at C source code level}. Programming in C is not sufficient for implementing separation kernels and programmers have to manipulate hardware directly by embedding assembly code in C. The assembly code is often omitted in full formal verification (e.g., seL4 \cite{Klein09}) and not supported by code abstraction tools, such as CParser \cite{GAK12} which translates a large subset of C-99 code into Isabelle/HOL. Existing works have to be extended for full formal verification considering C and assembly code together. 


\subsection{Dealing with Multicore and Concurrency}
In the domain of high-assurance systems, an increasing trend is the adoption of multicore processor to fulfil demands of higher computing power \cite{parkinson11}. The overall performance of systems is improved by concurrent execution of instructions in multicore processors. The latest version of ARINC 653 \cite{ARINC653p14} specifies the functionality and system services of multicore separation kernels. As summarized in {\tabprefix} \ref{tab:impls1}, separation kernels from industry and academia mostly support multicore processors. 
Multicore kernels are challenging formal verification and the safety/security certification \cite{Cohen13}. To the best of our knowledge, there is no research work on formal verification of multicore kernels in the literature. 

Separation kernels are reactive systems whose execution is triggered by system calls and in-kernel events. 
In general, the execution of system calls of monocore kernels are non-preemptive. It is often assumed in formal verification that kernels do not have in-kernel concurrency and the execution of functions handling events is considered to be atomic, such as in \cite{Klein09}. 
In such a case, formal verification of critical properties could be decomposed to examine individual execution steps, i.e., atomic functions. This is the basic idea of the unwinding theorem \cite{rushby92} to reason about noninterference. 
However, kernels are preemptive when processing other interruptions and thus in-kernel concurrency exists in practice. On the other hand, multicore introduces more complicated concurrency in separation kernels. The complexity increases greatly due to concurrent execution among cores and the shared resources. Functions to handle events are shared-variable based parallel programs and are executed in an interleaved manner. 

A promising way of conquering this issue is compositional verification \cite{Shankar93,Young13}. Rely-guarantee method \cite{Jones83} is a fundamental approach for compositional reasoning of parallel programs with shared variables. We outline the challenges and potential directions in formal methods application on multicore separation kernels as follows.

(1) \emph{Formalization of critical properties}. The original critical properties for separation kernels are usually defined on a state machine in which a transition is a big-step action (e.g., a system call). In the case of multicore, non-atomicity of events requires new formalization of critical properties. 

(2) \emph{Specification languages in theorem provers}. Existing specification of separation kernels uses inherent \emph{functions} of programming languages in theorem provers (e.g., Isabelle/HOL, Coq) to specify the atomic behavior of events. For multicore, specification languages which can express interleaved semantics and deal with complexity are required in theorem provers.

(3) \emph{Compositional reasoning of critical properties}. 
Although compositional reasoning of language-based information flow security has been studied \cite{Mantel11,Murray16}, compositional reasoning of state-event based definitions, which are usually applied on operating system kernels, should be addressed in future. Compositional reasoning of other critical properties also deserves further study. Proof systems for compositional reasoning and their automation techniques are critical. 

(4) \emph{Parallel refinement}. Based on the specification languages, a refinement framework is certainly needed with considerations of concurrency and compositionality of refinement relation \cite{LiangFF14}. The critical properties of separation kernels are necessary to be preserved during parallel refinement of multicore specification. 



\subsection{Formal Development and Code Generation}

\begin{table}
\tbl{Statistics of Challenges \label{tab:chall_comp}}{%
\begin{tabular}{|L{4.6cm}|C{0.15cm}|C{0.15cm}|C{0.15cm}|C{0.15cm}|C{0.15cm}|C{0.15cm}|C{0.15cm}|C{0.15cm}|C{0.15cm}|C{0.15cm}|C{0.15cm}|C{0.15cm}|C{0.15cm}|C{0.15cm}|C{0.15cm}|}
\hline
\centering
\multirow{2}{*}{\textbf{Related Work}} & \multicolumn{5}{c|}{\textbf{\tabincell{c}{Specification \\ Bottleneck}}} & \multicolumn{4}{c|}{\textbf{\tabincell{c}{Full Formal \\ Verification}}} & \multicolumn{4}{c|}{\textbf{\tabincell{c}{Multicore \\ Concurrency}}} & \multicolumn{2}{c|}{\textbf{\tabincell{c}{Formal \\ Dev.}}} 
\\ \cline{2-16}
 & 1 & 2 & 3 & 4 & 5 & 1 & 2 & 3 & 4 & 1 & 2 & 3 & 4 & 1 & 2
\\\hline \hline

GWV \cite{Greve03} & & $\almostaddress$ & $\almostaddress$ & & & & & & & & & & & &
\\\hline
Noninterference \cite{rushby92,von04,Murray12} & & $\almostaddress$ & $\almostaddress$ & & & & & & & & & & & &
\\\hline
MASK \cite{Martin00,Martin02} & & $\almostaddress$ & $\almostaddress$ & $\almostaddress$ & & & & & & & & & & $\almostaddress$ & $\almostaddress$
\\\hline
MPS \cite{AFBMRTO02} & & & $\almostaddress$ & $\almostaddress$ & & & & & & & & & & &
\\\hline
An ARINC Scheduler \cite{Singhoff07} & $\almostaddress$ & & & & & $\almostaddress$ & & & & & & & & &
\\\hline
Craig \cite{Craig07} & & & $\almostaddress$ & $\almostaddress$ & & & & & & & & & & & $\metioned$
\\\hline
LPSK \cite{Phelps08} & & $\metioned$ & $\almostaddress$ & $\almostaddress$ & & & & & & & & & & $\almostaddress$ & $\metioned$
\\\hline
SPK \cite{Andre09} & & & $\almostaddress$ & $\almostaddress$ & & & $\almostaddress$ & & & & & & & $\almostaddress$ &
\\\hline
OS-K \cite{Kawamorita10} & & & $\almostaddress$ & $\almostaddress$ & & & $\almostaddress$ & & & & & & & $\almostaddress$ &
\\\hline
Verified Software \cite{Velykis10} & & $\almostaddress$ & $\almostaddress$ & $\almostaddress$ & & & & & & & & & & &
\\\hline
Xenon \cite{freit11} & & & $\almostaddress$ & $\almostaddress$ & & & & & & & & & & &
\\\hline
CISK \cite{Verb14} & & & $\fulladdress$ & $\almostaddress$ & & & & & & & & & & &
\\\hline
ARINC 653 Standard \cite{zhao15} & & & $\almostaddress$ & $\almostaddress$ & & & $\almostaddress$ & $\almostaddress$ & & & & & & $\almostaddress$ &
\\\hline

ED \cite{Heitmeyer06,Heitmeyer08} & & $\metioned$ & $\almostaddress$ & $\almostaddress$ & & & & & $\almostaddress$ & & & & & & $\almostaddress$
\\\hline
AAMP7 \cite{Greve04,Wilding10} & & $\almostaddress$ & & & & & & & $\almostaddress$ & & & $\metioned$ & & & $\almostaddress$
\\\hline
INTEGRITY-178B \cite{Richards10} & & $\almostaddress$ & $\almostaddress$ & $\almostaddress$ & & & & & $\almostaddress$ & & & & & & $\almostaddress$
\\\hline
PikeOS \cite{Baumann11,Tverdy11,Bond14,Klaus15} & & $\almostaddress$ & $\almostaddress$ & $\almostaddress$ & & $\almostaddress$ & & & $\almostaddress$ & & & & & &
\\\hline
seL4 \cite{Murray13} & & $\almostaddress$ & $\almostaddress$ & $\almostaddress$ & & & & & $\fulladdress$ & & & & & &
\\\hline
PROSPER \cite{Dam13} & & & & & & & & & $\fulladdress$ & & & & & &
\\\hline
XtratuM \cite{sanan14} & & & $\almostaddress$ & $\almostaddress$ & & & & & & & & & & $\almostaddress$ &
\\\hline
mCertiKOS \cite{Costanzo16} & & & $\almostaddress$ & $\almostaddress$ & & & & & $\fulladdress$ & & & $\almostaddress$ & & $\almostaddress$ &
\\\hline
ARINC 653 \cite{Zhao16} & & $\almostaddress$ & $\almostaddress$ & $\almostaddress$ & & & & & & & & & & $\almostaddress$ &
\\\hline
DEOS \cite{Penix00,Penix05,Ha04} & $\almostaddress$ & & & & & $\almostaddress$ & & & $\almostaddress$ & & & & & &
\\\hline
A VxWorks scheduler \cite{Asberg11} & $\almostaddress$ & & & & & $\almostaddress$ & & & & & & & & &
\\\hline
RTSJ scheduler \cite{Zerzelidis06b,Zerzelidis10} & $\almostaddress$ & & & & & $\almostaddress$ & & & & & & & & &
\\\hline

\end{tabular}}
\begin{tabnote}%
\Note{$\fulladdress$:}{the challenge has been addressed} 
\Note{$\almostaddress$:}{the challenge has been partially addressed} 
\Note{$\metioned$:}{ the authors have mentioned the challenge but failed to address it} 
\Note{}{The blank is that the literature does not mention this kind of problem.} 
\end{tabnote}%
\end{table}

Separation kernels are always formally verified by the post-hoc approach, i.e., formal verification on an existing implementation. 
One promise of formal methods is to develop formal models step by step and generate code automatically or manually from the model whose correctness and properties have been formally verified. The benefit of formal development for separation kernels is significant. First, the specification and the verification targets, i.e. implementations of separation kernel, are developed in tandem, the specification bottleneck can be greatly alleviated. Second, formal proofs requested by safety/security certification can be generated during refinement-based development. Third, developing source code is a time-consuming and error-prone process. Automatic code generation via certified/verified tools can alleviate many efforts to design and implementation and provide rigorous arguments to validate the generated code. For this purpose, the following challenges need to be addressed in the future. 

(1) \emph{Stepwise refinement for formal development supporting multicore}. In formal verification of seL4 \cite{Klein09,Murray13} and ED \cite{Heitmeyer08}, refinement methods have been applied. Due to the post-hoc verification objective of these projects, refinement is not a technique to develop the specification in a stepwise manner, but to prove the conformance between formalizations at different levels. Therefore, they have few levels of specification and the refinement is coarse-grained. For the purpose of formal development, a stepwise refinement framework, which is able to deal with additional design elements (e.g., new events and new state variables) and concurrency, is highly desired. 

(2) \emph{Verified code generation and traceability}. Formal synthesis of separation kernels is difficult since the code should be very efficient and embedded with assembly code to manipulate hardware. Therefore, the machine model have to be considered in the formal synthesis. On the other hand, verified synthesis and traceability of the code to formal models are required for certifications. 


\subsection{Summary}

We have compared typical related work which have (partially) addressed the challenges and studied the potential directions mentioned in {\tabprefix} \ref{tab:chall_comp}. From the table, we could see that the challenge of specification bottleneck has been widely considered, in particular the generality and reusability of formal specification. Full formal verification has attracted large efforts in recent years, e.g., seL4, mCertiKOS, and PROSPER, in which full formal verification of the source code or binary code has been done. As a new trend in high-assurance systems, multicore and concurrency issues in formal verification of separation kernels have not been addressed. To the best of our knowledge, except some efforts to preemptive and interruptable OSs \cite{Chen16,Xu16}, there is no research work on formal verification of multicore kernels. Formal development and code generation for separation kernels has been partially considered in some research works. However, issues considering automatic code generation have not been addressed.

\section{Conclusion}
\label{sec:concl}
We have surveyed, categorized and comparatively analyzed major research works in formal methods application on separation kernels. Our analytical framework clarifies the scope of formal methods application on separation kernels and characterizes the separation kernels. The taxonomy and survey of research works have distilled existing efforts in this field to the current date. This survey additionally gives an overview and limitations of existing works by a detailed comparison and analysis. We also highlight the challenges and future directions. With this snapshot of the overall research landscape, we thus hope the separation kernel community can better explore various potential opportunities to further improve the safety and security of separation kernel implementations and reduce the cost of development and certification by formal methods application. 





\end{document}